\documentclass[aps,prd,twocolumn,superscriptaddress,nofootinbib,10pt]{revtex4-1}

\usepackage[utf8]{inputenc}
\usepackage{graphicx,amsmath,amsfonts,amssymb,aas_macros,slashed}
\usepackage{bbold}
\usepackage{datetime}
\usepackage{numprint}
\usepackage{enumitem}
\usepackage{graphicx,color,amsmath}
\usepackage{hyperref}

\usepackage{rotating}
\allowdisplaybreaks

\newcommand{\MeV}{\ensuremath{\mathrm{MeV}}}

\newcommand{\tmu}{{\tilde \mu}} 
\newcommand{\tE}{{\tilde E}}

\usepackage{pgfplots}
\usepackage{bm}
\usepackage{mathtools}

\makeatletter
\newcommand{\thickhline}{%
    \noalign {\ifnum 0=`}\fi \hrule height 1pt
    \futurelet \reserved@a \@xhline
}
\newcolumntype{"}{@{\hskip\tabcolsep\vrule width 1pt\hskip\tabcolsep}}
\makeatother

\begin{document}

\title{Towards a full description of MeV dark matter decoupling:\\ 
a self-consistent determination of relic abundance and \boldmath$N_{\rm eff}$
}

\author{Xiaoyong Chu}
\email{xiaoyong.chu@oeaw.ac.at}
\affiliation{Institute of High Energy Physics, Austrian Academy of Sciences, Georg-Coch-Platz 2, 1010 Vienna, Austria}
\author{Jui-Lin Kuo}
\email{juilink1@uci.edu}
\affiliation{Department of Physics and Astronomy,
University of California, Irvine, CA 92697-4575, USA}
\author{Josef Pradler}
\email{josef.pradler@oeaw.ac.at}
\affiliation{Institute of High Energy Physics, Austrian Academy of Sciences, Georg-Coch-Platz 2, 1010 Vienna, Austria}
\affiliation{CERN, Theoretical Physics Department, 1211 Geneva 23, Switzerland}

\begin{abstract}
Thermal dark matter at the MeV mass-scale has its abundance set during the highly non-trivial epochs of neutrino decoupling and electron annihilation. The technical obstacles attached to solving Boltzmann equations of multiple interacting sectors being both relativistic and non-relativistic have to-date prevented the full treatment of this problem. Here, for the first time, we calculate the freeze-out of light dark matter, taking into account the energy transfer between the dark sector, neutrinos, and the electromagnetically interacting plasma from annihilation {\it and} elastic scattering processes alike. We develop a numerically feasible treatment that allows to track photon and neutrino temperatures across freeze-out and to arrive at a precision prediction of $N_{\rm eff}$ for arbitrary branching ratios of the dark matter annihilation channels. In addition, our treatment resolves for the first time the dark matter temperature evolution across freeze-out involving three sectors. It enters in the efficiency of velocity-dependent annihilation channels and for a flavor-blind $p$-wave annihilation 
into electron- and neutrino-pairs of all generations, we find the present Planck data excludes  a complex scalar dark matter particle of mass of $m_\phi \leq 7$~MeV.
\end{abstract}

\maketitle

\section{Introduction}

Electroweak-scale dark matter (DM) which undergoes thermal freeze-out enjoys a great  advantage for the practitioner:  its abundance is set at a time when it typically leaves no trace in the following observationally accessible epochs of neutrino decoupling and big bang nucleosynthesis (BBN). The story of the Universe is hence one that evolves in sequence where DM genesis is relegated into the deeper radiation dominated phases.
In contrast, when considering MeV-scale thermal DM, it chemically decouples right during those highly-nontrivial later epochs. The annihilation of DM then affects cosmological observables, most notably the number of relativistic degrees of freedom~$N_{\rm eff}$. 
In fact, standard lore has it that a thermal dark state with a mass at or just above MeV is excluded as its annihilation heats  the photon and neutrino baths unequally and induces a change in their temperatures
that pertains to the well constrained epoch of recombination.

The calculation of $N_{\rm eff}$ from Standard Model (SM) physics alone has a long history~\cite{Dolgov:2002wy}, and the greatest of efforts have been poured into making its prediction as precise as possible~\cite{Mangano:2005cc,deSalas:2016ztq}.
Systematic treatments of light MeV-scale DM decoupling and with it a {\it joint} determination of $N_{\rm eff}$ have a likewise rich history, but are  comparatively less technical in their demand. In fact, most available works assume instantaneous
neutrino decoupling~\cite{Ho:2012ug,Boehm:2012gr,Steigman:2013yua,Boehm:2013jpa,Green:2017ybv}; related works that also consider the modifications of light element abundances during primordial nucleosynthesis are~\cite{Serpico:2004nm,Nollett:2013pwa,Nollett:2014lwa,Kawasaki:2015yya,Wilkinson:2016gsy,Depta:2019lbe,Berlin:2019pbq,Sabti:2019mhn,Berlin:2019pbq,Sabti:2021reh,Giovanetti:2021izc}. Only recently were dedicated efforts towards a systematic treatment of MeV-scale thermal DM decoupling with a precision determination of $N_{\rm eff}$  made in~\cite{Escudero:2018mvt} as well as in~\cite{Depta:2019lbe, Sabti:2019mhn}.
These works put the SM processes and DM decoupling on the same footing and include the effects of energy transfer between the sectors from annihilation of DM into the observable sector.

This is, however, not the final answer. First, previous works had to assume that DM stays in thermal equilibrium with either photons or neutrinos placing a principal restriction on the relative branching into electrons/photons and neutrinos and dominance of one over the other had to be assumed. Second, energy transfer due to {\it elastic} scattering processes were not included. Since the rate of elastic collisions dominates over the annihilation rate in non-relativistic freeze-out, this leaves a lingering doubt on the theoretical uncertainty of the $N_{\rm eff}$ prediction.
Moreover, elastic collisions enter the prediction of the DM temperature, with the latter feeding into the annihilation efficiency. 
Therefore, in order to obtain both  $N_{\rm eff}$ and the final DM abundance self-consistently, we must extend the scope of previous works. 
For this undertaking, one solves a system of coupled Boltzmann equations over a great dynamical range that spans from relativistic to non-relativistic regimes for DM and electrons, with rates that can exceed the Hubble rate by many orders of magnitude, resulting in excruciating numerical demand on precision when detailed balancing conditions have to be fulfilled. 

In this work we overcome those obstacles and, for the first time, present a complete treatment of MeV-scale DM decoupling. 
We take into account all relevant two-to-two scattering processes --- inelastic and elastic. We allow for individual temperatures in electrons/photons, neutrinos, and in the dark sector and for the latter two include the chemical potential. 
By making some minimal assumptions on the particle distribution functions, we are able to cast the problem of the three coupled sectors into a form that is amenable to numerical solution. It affords us a solution of the DM relic abundance with arbitrary branching ratios into electron/photons and neutrinos and at the same time provides a prediction of $N_{\rm eff}$, measured from the cosmic microwave background (CMB) data, that includes all apparent dominant contributions. 
The purpose of this work is to lay out the methodology and demonstrate it on an exemplary complex scalar DM model with flavor-blind $Z'$ mediated couplings to charged and neutral leptons. It is to be followed up by considering several classes of MeV-scale new physics~\cite{companion}.

The paper is organized as follows. In Sec.~\ref{sec:overview} we provide an overview over the coupled sectors system. In Sec.~\ref{sec:approxtherm} the approximations that we take on the thermodynamics are discussed. In Sec.~\ref{sec:collision_integrals} this is followed up by detailing the associated collision terms. Our results and numerical solution are presented in Sec.~\ref{sec:numsolution} and the cosmological $N_{\rm eff}$ constraint is discussed in Sec.~\ref{sec:deltaNeff}. Conclusions are presented in Sec.~\ref{sec:conclusions}, followed by several appendices providing additional explicit expressions that enter our calculations.

\section{Three-sector system overview}
\label{sec:overview}

 \begin{figure*}[t]
\begin{center}
\includegraphics[width=\textwidth,viewport=0 650 1470 1080,clip=true]{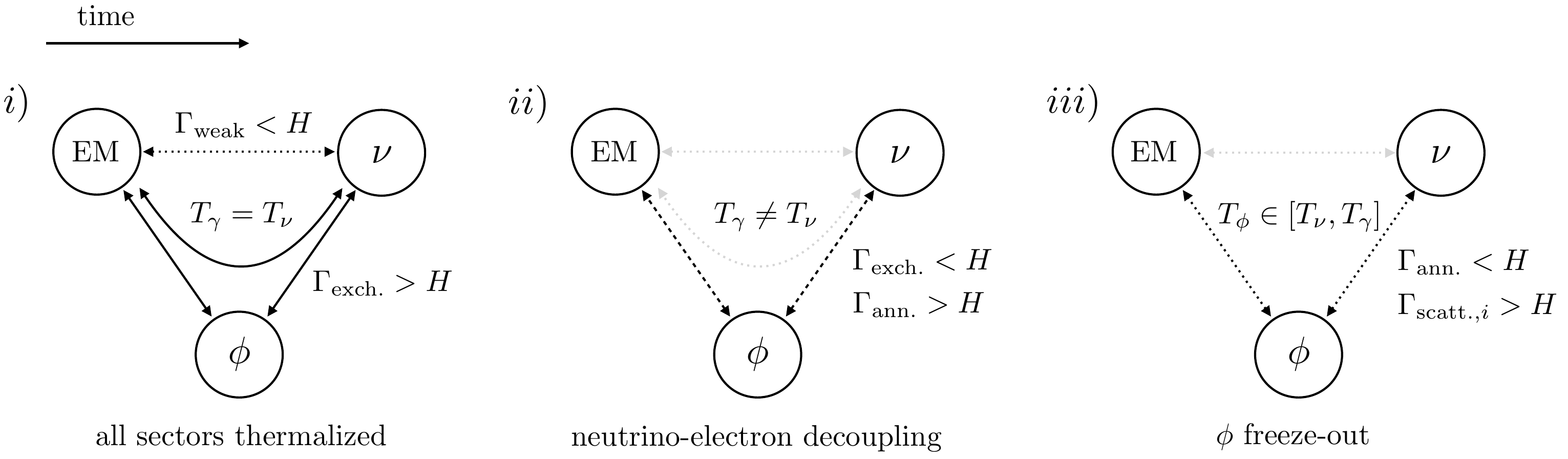}
\caption{Schematic depiction of an exemplary evolution of the the three coupled sectors (EM, $\nu$, $\phi$) in time. {\it Left panel:} at an early stage, even when SM weak interactions have decoupled, $\Gamma_{\rm weak} < H$, energy exchange processes with rate $\Gamma_{\rm exch.}>H$ can keep the EM and $\nu$ sectors equilibrated and $T_\nu = T_\gamma$ follows. {\it Middle panel:} once energy exchange becomes inefficient, $\Gamma_{\rm exch.}<H$, the neutrino and EM sectors are separately heated by the respective by annihilation processes with $\Gamma_{\rm ann.}>H$ and photon and neutrino temperatures evolve separately, $T_\nu\neq T_\gamma$ follows. {\it Right panel:} DM freezes out when $\Gamma_{\rm ann.} < H$ but the dark sector temperature $T_\phi$ continues to evolve due to elastic scattering processes $\Gamma_{{\rm scatt.,}i}>H$. Finally, when the latter rates become sub-Hubble, DM continues to cool adiabatically.
}
\label{fig:scheme}
\end{center}
\end{figure*}

We start our investigation by highlighting the essential stages in the time evolution of thermal sub-GeV DM freeze-out. The particle content of the Universe in the relevant epoch prior to/during BBN is given by 
the electromagnetically (``EM'') interacting particles (electrons, positrons and photons),  the active three flavors of SM neutrinos (``$\nu$''), and by the particles of the dark sector (``$\phi$''), where we assume that only the DM particle is populated in abundance and other particles associated with it have largely disappeared. For the sake of a comparative discussion, we  define the following thermally averaged interaction rates:
\begin{subequations}
\label{rates}
\begin{align}
    \Gamma_{\rm weak} & \equiv  n_e G_F^2 T_\gamma^2\,,      \\
    \Gamma_{\rm ann.} & \equiv n_\phi \langle \sigma_{\rm ann.} v \rangle \,,
    \\
    \Gamma_{{\rm exch.},i} & \equiv n_\phi^2  \langle \sigma_{{\rm ann.}, i} v \delta E\rangle /\rho_{i}\,
, \\
    \Gamma_{{\rm scatt.},i} & \equiv n_i \langle \sigma_{\rm scatt.}^{\phi i} v \rangle \, .
\end{align}
\end{subequations}
Here, $\Gamma_{\rm weak}$ informs us on the overall rate of SM weak interactions such as $ \nu e \leftrightarrow \nu e $. We anchor the rate on the electron number density $n_e$ to account for the latter's suppression once the photon temperature $T_\gamma$ drops below one~MeV; $G_F$ is the Fermi constant. The rate of DM annihilation $\Gamma_{\rm ann.} $ is given by the total thermally averaged annihilation cross section $\langle \sigma_{\rm ann.} v \rangle$ into all channels times the DM number density $n_\phi$. For the rate of energy exchange,  $\Gamma_{{\rm exch.},i}$ of DM with the EM $(i=e\text{~and~}\gamma)$ or neutrino  ($i=\nu$) sector, one additionally weighs the annihilation cross section by the injected energy. We define the rate normalized to the  energy density $\rho_i$. Finally, the rate of elastic DM scattering on target $i$ is obtained from the elastic scattering cross section $\sigma_{\rm scatt.}^{\phi i} $. All these rates are then to be compared to the Hubble rate~$H$. For example, when $\Gamma_{{\rm exch.},\nu}\sim H$ it means that the equivalent of the neutrino energy density between the $\phi$ and  neutrino sectors is exchanged in any volume of space within a Hubble time~$H^{-1}$. In this work, we shall assume similar efficiency of annihilation into the EM and $\nu$ sectors, so that we do not further keep track on the index $i$ on the exchange rate and simply write $\Gamma_{\rm exch.}$ instead. As we shall see explicitly below, $\Gamma_{{\rm exch.}}/H$ determines  when the departure from a common temperature shared by three sectors happens, while  $\Gamma_{\rm ann.}/H$ and $\Gamma_{{\rm scatt.},i}/H$ determine the DM freeze-out and kinetic decoupling temperatures from sector~$i$, respectively.

The sequence of decoupling for MeV DM is different than for weakly interacting massive particles (WIMPs). WIMP DM in thermal equilibrium with the SM bath  freezes out at a temperature $m_{\rm WIMP}/20 \gg 1$~MeV when the EM and $\nu$ sectors remain in perfect thermal equilibrium. In contrast, MeV DM typically decouples {\it after} weak interactions cease to keep the EM and $\nu$ sector thermally coupled. In other words, a fair fraction of energy density relative to SM  radiation can still be stored in the dark sector at the time of neutrino decoupling. 
The latter happens once~$\Gamma_{\rm weak} < H $. 

In the case of interest here, DM annihilation into both $\nu$ and EM sectors can then act as the agent that keeps the latter two sectors in equilibrium despite $ \Gamma_{\rm weak} < H$. This happens by virtue of energy exchange with associated rate $\Gamma_{\rm exch.}$. It provides a bridge between the EM and $\nu$ sectors and for as long as $ \Gamma_{\rm exch.} > H $, the temperatures of both sectors equilibrate, $T_\nu = T_\gamma$. This is depicted by the leftmost diagram \textit{i)} in Fig.~\ref{fig:scheme}. 

At some point, energy exchange will become less efficient, $\Gamma_{\rm exch.}< H$, leading to a departure of neutrino and photon temperatures, $T_\nu \neq T_\gamma$. Nevertheless, DM annihilation may still heat both sectors by annihilation into electrons, photons and neutrinos  as long as $\Gamma_{\rm ann.}>H$.
Evidently, neutrino and photon temperatures enter the  prediction of $N_{\rm eff}$ and it is one of the key objectives of the paper to track the relative evolution of $T_\gamma$ and $T_\nu$ to great precision. This stage is shown in the middle diagram \textit{ii)} of Fig.~\ref{fig:scheme}. 

Finally, DM chemically decouples and the comoving DM number freezes out when $\Gamma_{\rm ann.}< H$. 
The DM temperature itself, however, continues to evolve. It is determined by the more efficient elastic scattering processes $\Gamma_{{\rm scatt.}, i}$ with either the EM sector or with neutrinos. The canonical case here is that the DM temperature $T_\phi$ will be remain in the interval $T_\phi \in [T_\nu, T_\gamma] $. If kinetic equilibrium between $\phi$ and all its annihilation products is not maintained, the evolution of $T_\phi$ in fact enters the DM abundance prediction as demonstrated below. Here we show this effect for the first time explicitly. The DM freeze-out stage is schematically shown in the rightmost diagram \textit{iii)} of Fig.~\ref{fig:scheme}. 

Solving the gradual decoupling of three sectors in generality is a very demanding task because of the great dynamical range that enters in number- and energy-densities as well as in interaction rates. To our knowledge such treatment is not available in the literature to-date. In the remainder of the paper we  start from full generality of the system of coupled Boltzmann equations and deduce a numerically feasible treatment that allows to track all the three sectors with great accuracy. It involves a suitable approximate treatment of thermodynamic quantities and the specification of the integrated collision terms.

\section{Approximated Thermodynamics}
\label{sec:approxtherm}

The fundamental quantity describing a particle species in a homogeneous and isotropic expanding Universe is the (spatially averaged) momentum distribution function $f(t, |\vec p|)$ where $t$ is the cosmic time and $|\vec p|$ is the magnitude of the particle's three momentum. Its time evolution is governed by the Boltzmann equation
\begin{align}
\label{eq:boltzmann}
   \dfrac{\partial f}{\partial t} - H \dfrac{|\vec p|^2}{E} \dfrac{\partial f}{\partial E} = \dfrac{1}{E} C[f]\,,
\end{align}
where $H$ is the Hubble rate and $E$ is the energy associated with $|\vec p|$. The collision integral $C[f]$ accounts for  particle interactions. In this work, we are chiefly concerned with two-body  processes $p_1 + p_2 \leftrightarrow p_3 + p_4$, where $p_i $ denote four-momenta.  The collision integral of species ``1'' with a momentum distribution in $p_1$ is then given by 
\begin{align}
\label{eq:collisionterm}
    C[f_1] &= - \dfrac{S g_2}{2} \int d\Pi_{i = 2,3,4}\,(2\pi)^4 \delta^{(4)} (p_1 + p_2 - p_3 -p_4) \nonumber \\
    &\times J \dfrac{1}{g_1 g_2} \sum\limits_{\rm spins} |\mathcal{M}_{12\leftrightarrow 34}|^2\,.
\end{align}
Here, the factor $2$ in the denominator ensures energy-momentum conservation in each collision, $S$ is a symmetry factor, $d\Pi_i = d^3 p_i/[ (2\pi)^3 2E_i]$ is the Lorentz-invariant phase space element, 
and $|\mathcal{M}_{12\leftrightarrow 34}|^2$ is the squared matrix element of the scattering process in question.   Throughout the paper, $g_i$ stands for the $i$-particle degrees of freedom (without counting its antiparticle for non-self-conjugate species, and thus the same for $n_i$); see Appendices for the explicit values.\footnote{A factor of $1/2!$ additionally needs to be supplied on the right-hand-side of~\eqref{eq:collisionterm} for each pair of identical particles in initial or final states, while another factor of $2$ will be multiplied for two identical particles in the initial state.}  The quantum statistical factor $J$ that weighs each scattering is given by 
\begin{align}
\label{eq:Jfactor}
J =   f_1 f_2(1\pm f_3)(1\pm f_4) - (1\pm f_1)(1\pm f_2)f_3 f_4  \,, 
\end{align}
where the $+(-)$ sign applies to bosons (fermions) with the usual  interpretation of Bose-enhancement (Pauli-blocking).%
\footnote{The calculation is based on the assumptions that the process conserves time reversal, or, equivalently CP symmetry, and that inhomogeneities and anisotropies can be neglected.}

Integrating Eq.~\eqref{eq:boltzmann} with different weights over the phase space of the particle species of interest, we obtain for each sector the respective evolution of number and energy densities, $n_i$ and $\rho_i$, 
\begin{align}
       &  \dfrac{ \partial n_i}{\partial t} + 3Hn_i \equiv \dfrac{\delta n_i}{\delta t}\,, \label{eq:BZMn} \\
   & \dfrac{\partial  \rho_i}{\partial t} + 3H(\rho_i + P_i)\equiv \dfrac{\delta \rho_i}{\delta t}   \,,\label{eq:BZMrho}
\end{align}
where $P_i$ is the pressure density. Explicitly, the overall evolution is obtained by summing all contributing annihilation channels of the number densities,  and all two-body channels (both annihilation and scattering) for the energy densities,
\begin{align}
   & \dfrac{\delta n_i}{\delta t} =   g_i \int \dfrac{d^3 p_i}{(2\pi)^3 E_i}\,\sum_{\rm ann.} C[f_i]   \,, \\
   &  \dfrac{\delta \rho_i}{\delta t}  =   g_i \int \dfrac{d^3 p_i}{(2\pi)^3 E_i}\,  \, \sum_{\rm all} C[f_i] \,\delta E \,, 
\end{align}
 where $\delta E$ is the energy exchange for each channel.\footnote{If the corresponding channel $1+2 \to 3+4$ is   pair annihilation, there is $\delta E = E_1 +E_2 $, counting the contributions of both particle $1$ and antiparticle $2$. For elastic scattering,  $\delta E = E_3 -E_1$, which is the energy change for particle $1$($\equiv 3$); see below for details.} In summary, $\delta n_i/\delta t $ is the interaction rate of number-changing processes with another sector and $\delta \rho_i /\delta t$ is the energy exchange rate with another sector.

As stated before, the general (numerical) solution of such set of coupled Boltzmann equations~\eqref{eq:BZMn} and~\eqref{eq:BZMrho} over a great dynamic range is in practice unfeasible. For the three-sector system of our interest, we now show how  taking  some minimal approximations opens the path to a  numerical solution. The approximations fall into two categories: the ones to describe the momentum distribution functions and the ones to obtain detailed balance, which we shall discuss below in turn.

\subsection{Momentum distribution functions}

The aim of this subsection is to show that the evolving momentum distribution functions can be well described by the  evolution of the  temperature, $T_i$, and the chemical potential, $\mu_i$,   in each sector ($i= \gamma/e,\, \nu,\,\phi$). Therefore,  Eqs.~\eqref{eq:BZMn} and \eqref{eq:BZMrho}  can be replaced by the differential functions for $T_i$ and $\mu_i$ through the relations given in App.~\ref{sec:conversion}.

Prior to their decoupling, each sector is expected to maintain kinetic equilibrium at temperature $T_i$ so that their  momentum distribution functions are given by
\begin{equation}
    f_i (E_i,\mu_i) = \dfrac{1}{e^{(E_i- \mu_i)/T_i} \mp 1 }\,,\label{eq:generalMDF}
\end{equation}
where $\mu_i$ is the chemical potential of species $i$, and ``$-$ ($+$)'' applies to boson (fermion); the  distribution function in chemical equilibrium is denoted by 
$$f^{\rm eq}_i (E_i) \equiv  f_i (E_i,\mu_i =0)\,.$$
In the following, we elaborate on the further approximations taken on the respective sectors. To this end,  we use the tilde overscript to denote  dimensionless parameters normalized to their corresponding temperature, \textit{e.g.}, $\tE_i \equiv E_i/T_i$ and $\tmu_i \equiv \mu_i/T_i$.
 
\paragraph*{EM sector} The
EM sector contains the photon, electrons and positrons. Kinetic equilibrium ensures a common temperature $T_\gamma$. While the photon can be simply described by a black-body spectrum, the electron carries chemical potential $\mu_e$ induced by the baryon asymmetry of the Universe. However, before electron freeze-out $\mu_e$ is negligible, and once it becomes relevant the electron abundance is too diminished to affect the other two sectors any longer. Therefore, in our calculations we  set $\mu_e =0$ and take its  momentum distribution to be a thermal one:  
\begin{align}
\label{eq:dist_e}
    f_e (\tE_e) \simeq \dfrac{1}{e^{\tE_e} + 1 }\quad (T_\gamma \gtrsim m_e/20)\,.
\end{align}

\paragraph*{Neutrino sector}
For all processes of interest in this work, neutrinos can be treated as being massless. In a standard cosmological history, their distribution closely tracks a thermal one~\cite{Smith:2008ic}. Also in our case, we expect any departure from the Fermi-Dirac distribution (at non-zero chemical potential) for neutrinos to be small.\footnote{The elastic scattering rate per neutrino,  $n_\phi \langle \sigma_{\nu\phi \to \nu \phi} v \rangle$, is approximately of the same order as  the DM annihilation rate, $n_\phi \langle \sigma_{\phi  \phi \to \nu  \nu} v \rangle$, by virtue of crossing symmetry. As a result, when  DM annihilation is efficient, the re-distribution of  momenta among neutrinos  through $\nu + \phi \to \nu + \phi$ is expected to be  efficient too.  
{\it After} non-relativistic DM freeze-out, the process $\phi + \phi \to \nu + \nu$  dominates over its inverse, rendering the exact distribution of neutrinos less relevant.}
 Finally, we neglect the mild differences among the three flavors and adopt a flavor-blind momentum distribution function with a first-order expansion of Eq.~\eqref{eq:generalMDF} in the chemical potential~$\mu_\nu$,
\begin{align}
\label{eq:dist_nu}
	f_\nu (\tE_\nu,\tmu_\nu) %
 \simeq f^{\rm eq}_\nu ( \tE_\nu) + \tmu_\nu  f^{(1)}_\nu (\tE_\nu) + \mathcal{O}(\tmu_\nu^2) \,,
\end{align}
where 
\begin{align}
\label{eq:fnuone}
  f_\nu^{(1)}(\tE_\nu) = \frac{1}{e^{\tE_\nu }+e^{- \tE_\nu} + 2}\,.
\end{align}
The expansion is justified since neutrinos freeze out relativistically and are always well populated such that $\tmu_\nu \ll 1$. We will not consider any contributions at $\mathcal{O}(\tmu_\nu^2)$ order. 

\paragraph*{Dark sector} In this work, we focus on the canonical case where light DM stays in kinetic equilibrium with itself until it is decoupled from the EM and neutrino sectors. The DM  distribution function is then characterized by a dark-sector temperature $T_\phi$  and a chemical potential $\mu_\phi$.%
\footnote{Such assumption works well for DM freeze-out in two-sector systems~\cite{Binder:2021bmg} and  is expected to hold for DM with  MeV mass. In the context of self-interacting DM~\cite{Spergel:1999mh}, MeV DM may stay kinetically self-coupled until $T_\phi\sim {\rm eV}$~\cite{Kamada:2017gfc}.}
Considering a symmetric DM state and non-relativistic freeze-out, its momentum distribution function is described by $ ({e^{\tE_\phi} \mp 1 })^{-1}$ well before freeze-out, and by $e^{-(\tE_\phi -\tmu_\phi )}$ at freeze-out and later.  Both expressions can be unified by the approximate form
\begin{align}
\label{eq:dist_massive}
	 	f_\phi (\tE_\phi, \tmu_\phi) \simeq  \frac{e^{\tmu_\phi}}{e^{\tE_\phi } \mp 1} =    e^{ \tmu_\phi } f^{\rm eq}_\phi (\tE_\phi)\,.
\end{align}
Its integrated form yields the number density as  $n_\phi  =   e^{\tmu_\phi} {n_\phi^{\rm eq}(T_\phi)}$.

\subsection{Statistical factors in annihilation and scattering}

Above we have discussed the assumptions for factoring out the chemical potential for each momentum distribution function. In this subsection, we introduce the assumptions that allow us to do the same for the quantum statistic factor of each process. 

We first re-write the statistic factor $J$ in Eq.~\eqref{eq:Jfactor} in a form that collects the difference of the forward and inverse process as,  
\begin{align}
 f_1 f_2 (1 \pm f_3) (1\pm f_4) \left( 1 - e^{\tfrac{E_1 -\mu_1 }{T_1}+\tfrac{E_2 -\mu_2 }{T_2} -\tfrac{E_3 -\mu_3 }{T_3} -\tfrac{E_4 -\mu_4 }{T_4} } \right)\,,\notag
\end{align}
where in each factor (+) applies to a boson and ($-$) to a fermion in the respective final state.  If all sectors are thermalized with equal temperature, the detailed balance condition $J=0$ is manifest by observing that $E_1  +E_2  = E_3  + E_4  $ and $\mu_1 +\mu_2 = \mu_3 + \mu_4 $; importantly detailed balance is attained independent of the approximations for $f_i$ introduced above. 
In the numerical evaluation, the above parametrization is crucial to maintain detailed balance when the associated interaction rates become much larger than the Hubble rate. 
Finally, to further simplify the expression of $J$ as a function of $\mu_i$,  we  neglect the quantum statistical factors in the final states, $1\pm f_{3,4} \simeq 1$. The uncertainties associated with it are studied in the following subsection.\footnote{In our treatment one may take into account below the sub-leading contributions from final state statistics, in terms of $f_1 f_2f_{3,4}$, on the right hand side of Eqs.~(\ref{eq:Jfactor_ann}-\ref{Jscatt}) in a straightforward manner.}

For the purpose of this work, we consider only symmetric abundances of DM. For particle annihilation $1+ 2 \leftrightarrow 3 + 4$, we  have $T_1 = T_2$ ($\mu_1 = \mu_2$) and $T_3 = T_4$ ($\mu_3 = \mu_4$), together with energy conservation $E_1 +E_2 = E_3 + E_4 \equiv E_+$.  We now choose the convention that the final state is the {\it heavier} particle-pair involved. This way, when we neglect the Boltzmann-suppressed final state quantum statistical factors as discussed above, we still retain the more important initial state quantum statistical factors of neutrinos and electrons in the DM annihilation process. This allows us to write,
\begin{align}
  J & = f_1 f_2 (1\pm f_3)(1\pm f_4) \left[ 1- e^{ (T_1^{-1} - T_3^{-1}) E_+} e^{2(\tmu_3 - \tmu_1)}\right] \nonumber \\
  &\simeq f_1 f_2  -  f_1 f_2 \Delta_{\rm ann.}   \beta_{\rm ann.} \nonumber \\
  \label{eq:Jfactor_ann}
  & =  f_1 f_2 (1-\Delta_{\rm ann.})  + f_1 f_2 \Delta_{\rm ann.} (1- \beta_{\rm ann.})\,,
\end{align}
where $\Delta_{\rm ann.} \equiv e^{ (T_1^{-1} - T_3^{-1}) E_+}$ and $\beta_{\rm ann.} \equiv e^{2(\tmu_3 - \tmu_1)} =  e^{2(\tmu_4- \tmu_2)}$. In the last line of this equation, the first term vanishes when two sectors are in kinetic equilibrium, \textit{i.e.}, when $T_1 = T_3$, while the second term only vanishes when  thermal equilibrium is reached,  \textit{i.e.}, when $\tmu_1 = \tmu_3$.

Elastic scattering is another important element of this paper that, for the first time, makes the calculation of  energy transfer among the sectors self-consistent.
Here, the temperatures and chemical potentials of the respective scattering states remain unaltered. 
By appropriate assignment we set $T_{1(2)} = T_{3(4)}$ as well as $\mu_{1(2)} = \mu_{3(4)}$. In the integration of the corresponding collision terms, $J$ is to be weighted by the energy transfer per scattering, $\delta E \equiv E_3 - E_1 = E_2 -E_4$. Notice that   $J$ counts both the process and its inverse, which are the same in the case of elastic scattering, 
so an additional factor of $1/2$ is added below to avoid double counting after integrating over all possible initial momenta.
Similarly, after again neglecting the statistical factor  for the final states, $1\pm f_{3,4} \simeq 1$, we obtain %
\begin{align}
 \dfrac{1}{2} \, J   &= \dfrac{1}{2} \, f_1 f_2 (1\pm f_3)(1\pm f_4) \left[ 1- e^{(T_2^{-1}-T_1^{-1})\delta E}\right]  \nonumber \\
  \label{Jscatt}
  &\simeq f_1 f_2 \Delta_{\rm scatt.}\,,
\end{align}
with $\Delta_{\rm scatt.} = [1- e^{(  T_2^{-1} -  T_1^{-1}  )\delta E}] /2$. We observe, that once both sectors equilibrate in temperature, $T_1 = T_2$, $\Delta_{\rm scatt.}$ vanishes and there is no further energy transfer by elastic scattering processes. Evidently, there is no particle number change in the elastic scattering.

To summarize, we have assigned to each of the three sectors its own temperature and chemical potential, and subsequently introduced several approximations,  \eqref{eq:dist_e}--\eqref{Jscatt}. They allow  to treat the chemical potentials as pre-factors for  the distribution functions involved in the $2\to2$ processes. In addition, our formulation ensures that detailed balance is achieved without the cost of extremely high numerical precision. Next we  turn to the collision integrals, and show how the $T_i$ and $\mu_i$ variables can be separated in such a treatment.

\section{Collision Integrals}
\label{sec:collision_integrals}

In this section, we provide the expressions for the  collision terms in \eqref{eq:BZMn} and \eqref{eq:BZMrho}, and demonstrate that the chemical potentials $\tmu_i$ can be factorized, using~\eqref{eq:dist_nu} and~\eqref{eq:dist_massive}.

Concretely, the parametrization introduced above allows us to decompose the collision integrals into functions of chemical potentials and functions of temperatures,
\begin{align}
\label{dn}
    \dfrac{\delta n_i}{\delta t} &= \sum\limits_{i\neq j} a_{ij} \,\beta_{ij}(\tmu_i, \tmu_j) \,\gamma_{ij} (T_i, T_j) \,, \\
    \label{drho}
     \dfrac{\delta \rho_i}{\delta t} &= \sum\limits_{i\neq j} b_{ij} \,\beta_{ij}(\tmu_i, \tmu_j) \,\zeta_{ij} (T_i, T_j) \,,
\end{align}
where $a_{ij}$ and $b_{ij}$ are either $+1$ or $-1$, depending on the process, 
and $\beta_{ij}$ are combinations of the initial state chemical potentials such as $e^{\tmu_i+\tmu_j}$, $e^{\tmu_i+\tmu_j}(1-\beta_{\rm ann.})$, or  $\tmu_i e^{\tmu_j}$, normalized to  unity or zero when all chemical potentials vanish. 
The expressions for $\gamma_{ij} $ and $\zeta_{ij} $ are detailed below.  Note that elastic scattering processes only enter in $\zeta$, since particle number is conserved.
We also collect the analytical formulas for $\beta$, $\gamma$ and $\zeta$ for every interaction considered in this work in App.~\ref{app:collision}.

The advantage of such decomposition is that for each value of $m_\phi$ one may numerically tabulate the  functions $\gamma_{ij} $ and $\zeta_{ij} $ and avoid re-evaluating the time-consuming multiple integrals in each time step in the solution of the Boltzmann equations, or for each coupling strength of interaction. Moreover, one may readily improve the precision of our treatment, by adding tabulations of sub-leading corrections, such as working to $\mu_i^2$ order or including the quantum statistical factors for final state particles.

\subsection{Collision term for annihilation}

In the case of annihilation, particle 1\,(3) is the antiparticle of 2\,(4).
The simplified form of $J$ in~\eqref{eq:Jfactor_ann} for the annihilation process is independent of $E_{3,4}$. Therefore, we can simplify the collision integral by integrating over $d\Pi_{i=3,4}$, yielding the cross section $\sigma_{12\rightarrow 34}$.
In zeroth order of the chemical potentials, $\beta_{ij} = 1$, and the  collision term reads
\begin{align}
\label{eq:gamma12}
\gamma^{(0)}_{12\leftrightarrow 34} &= \dfrac{g_1 g_2}{(2\pi)^4} \int \dfrac{dsdE_+ dE_-}{2}\,f_1^{\rm eq} f_2^{\rm eq}\sigma_{12\rightarrow 34} \mathcal{F}_{12} \nonumber \\
&\times \left[ (1-\Delta_{\rm ann.}) + \Delta_{\rm ann.} (1- \beta_{\rm ann.})\right]\,,
\end{align}
where $s= (p_1+p_2)^2$ is the squared center-of-mass (CM) energy and $E_\pm = E_1 \pm E_2$. Note that the expression for $\gamma$ does not include the number change of anti-particles.  The flux factor reads,
\begin{align*}
\mathcal{F}_{12} = \sqrt{(p_1 \cdot p_2)^2 -m_1^2 m_2^2}= \dfrac{\sqrt{\lambda(s,m_1^2,m_2^2)}}{2}\,,
\end{align*}
with $\lambda(a,b,c) = a^2 +b^2 +c^2 -2 (ab+ac+bc)$ being the triangle function.
The Lorentz-invariant cross section, averaged over initial degrees of freedom, is given by~\cite{Gondolo:1990dk}
\begin{align}
\sigma_{12\rightarrow 34}  &= \dfrac{1}{4  g_1 g_2 \mathcal{F}_{12} } \int d\Pi_{i=3,4}\, (2\pi)^4  \nonumber \\
 &\times \delta^{(4)}(p_1 + p_2 - p_3 -p_4) \sum\limits_{\rm spins} |\mathcal{M}_{12\leftrightarrow 34}|^2\,.
\end{align}
One notes the relation $ g_1 g_2 \mathcal{F}_{12} \sigma_{12\rightarrow 34} = g_3 g_3 \mathcal{F}_{34} \sigma_{34\rightarrow 12}$. %
The kinematic ranges for the integration variables are
\begin{align*}
&s \geq \max \{(m_1 +m_2)^2, (m_3 + m_4)^2 \}\,, \quad E_+ \geq \sqrt{s}\,, \\
&E_- - E_+ \left( \dfrac{m_1^2 -m_2^2}{s} \right) \leq \left| \dfrac{2\mathcal{F}_{12}}{s} \sqrt{E_+^2 -s} \right|\,.
\end{align*}

The calculation of the energy-transfer collision terms for annihilation, 
$\zeta^{(0)}_{12\leftrightarrow34}$, is similarly obtained by supplying the energy-transfer factor ${\delta E = E_+}$ to the right hand side of Eq.~\eqref{eq:gamma12}.

\subsection{Collision term for elastic scattering}

For scattering, particle 1\,(2) is the same as 3\,(4), and thus $\gamma_{ij} \equiv 0$. 
To account for the energy transferred in elastic scattering processes, we need to introduce the additional Lorentz invariant variable $t = (p_1 -p_3)^2$, due to the $E_3$ dependence in~$\delta E$.
In this case, the $\zeta$ collision term at zeroth order in the chemical potentials can be expressed in terms of the differential cross section as 
\begin{align}
\label{eq:zeta12}
  \zeta^{(0)}_{12\leftrightarrow 12} &= \dfrac{g_1 g_2}{(2\pi)^4}\int dE_1 dE_2 ds dt\, f_1^{\rm eq} f_2^{\rm eq} \dfrac{d\sigma_{12\rightarrow 12}}{dt}  \nonumber \\ &\times\mathcal{F}_{12} \langle \Delta_{\rm scatt.}\delta E  \rangle\,,
\end{align}
where $\langle \Delta_{\rm scatt.} \delta E \rangle$ is the  energy transfer per scattering averaged over the azimuthal angle  $\phi^*$ in the CM frame.  The integration region of $t$ is given by $\left[ -\lambda(s,m_1^2,m_2^2)/s, 0 \right]$.

Solving the kinematics of the two-body scattering  in the medium (``lab'') frame, where $E_{1,2}$ are defined,  allows to separate $\delta E$ into $\phi^*$-independent and -dependent terms, $\delta E_{\rm scatt.}^0$ and $\delta E_{\rm scatt.}^1$,  given by 
\begin{align*}
  &\delta E_{\rm scatt.}^0 = \dfrac{(E_1 - E_2) st - (E_1 + E_2)(m_1^2 -m_2^2)t}{\lambda(s,m_1^2,m_2^2)} \,, \nonumber \\
  &\delta  E_{\rm scatt.}^1 = \dfrac{\sqrt{\lambda(s,m_1^2,m_2^2) t + st^2}}{\lambda(s,m_1^2,m_2^2)}  \cos\phi^* \nonumber \\
  &\times \sqrt{\lambda(s,m_1^2, m_2^2) + 4(m_1^2 E_2 + m_2^2 E_1) (E_1 + E_2) -4E_1 E_2 s}\,.
\end{align*}
 When considering the case that an initial state is at rest, $E_1 = m_1$, we obtain $\delta E_{\rm scatt.}^0  = -t/(2m_1)$ and $\delta E_{\rm scatt.}^1  = 0$, which gives the expected energy transfer per scattering in the lab frame.

After averaging over $\phi^*$ from $0$ to $2\pi$, we obtain 
\begin{widetext}
\begin{align}
  \langle \Delta_{\rm scatt.}  \delta E  \rangle = \left. \dfrac{\delta E_{\rm scatt.}^0}{2} - \dfrac{e^{(T_2^{-1} - T_1^{-1})\delta E_{\rm scatt.}^0}}{2}\left[\delta E_{\rm scatt.}^0 I_0\left(\dfrac{\delta E_{\rm scatt.}^1}{T_2} - \dfrac{\delta E_{\rm scatt.}^1}{T_1} \right) +  \delta E_{\rm scatt.}^1 I_1\left(\dfrac{\delta E_{\rm scatt.}^1}{T_2} - \dfrac{\delta E_{\rm scatt.}^1}{T_1} \right)\right]\right|_{\phi^* = 0}\,,
\end{align}
\end{widetext}
where $I_n$ is the modified Bessel functions of the first kind.
 
\subsection{Low-temperature regime}

So far, we have provided the formul\ae\ of factored collision integrals with thermal distribution functions and that are to be weighted with factors $e^{\tmu_i+\tmu_j}$.  As mentioned earlier, for processes involving neutrinos with distribution function $f_\nu$ we also include the second-order contribution $ \tmu_\nu f_\nu^{(1)}$ where $f_\nu^{(1)}$ is given in~\eqref{eq:fnuone}. At the time of neutrino decoupling $ \tmu_\nu$ is negligible. At low temperatures, however, the second-order term can become increasingly important as neutrinos develop a non-vanishing chemical potential. Thereby, $\tmu_\nu $ contributes in the prediction of $N_{\rm eff}$; see Eq.~\eqref{eq:nuenergyden}.
The expressions of the associated collision integrals are readily obtained by replacing $f_i^{\rm eq}$ with $f_i^{(1)}$ in the expressions for $\gamma$~\eqref{eq:gamma12} and $\zeta$~\eqref{eq:zeta12} above. 
 
Additionally, when  the dark state sector becomes deeply non-relativistic, extrapolations can be adopted to obtain the values of $\beta$ and $\gamma$, instead of  computing the integrals down to very small temperatures.\footnote{Uncontrolled errors typically  occur in numerical software when evaluating very small exponents.}  
Once $T_\phi \ll m_\phi$,  we typically also encounter $T_\phi \ll T_{\nu,\gamma}$. The latter condition becomes quickly fulfilled due to the adiabatic cooling, $T_\phi \propto 1/a^2$, of a decoupled non-relativistic species. Under such conditions, with $m_\phi/T_\phi \ge 100$,   we scale the ($p$-wave) interaction rates according to the temperature dependence of their non-relativistic thermal average. Thereby, one only integrates over the momentum distribution of the EM or neutrino sector. %
The  impact on the final value of $N_\text{eff}$ is  in practice negligible.  

\subsection{Discussion on introduced uncertainties}
\label{sec:errors}

 \begin{figure}[!t]
\begin{center}
\includegraphics[width=0.48\textwidth]{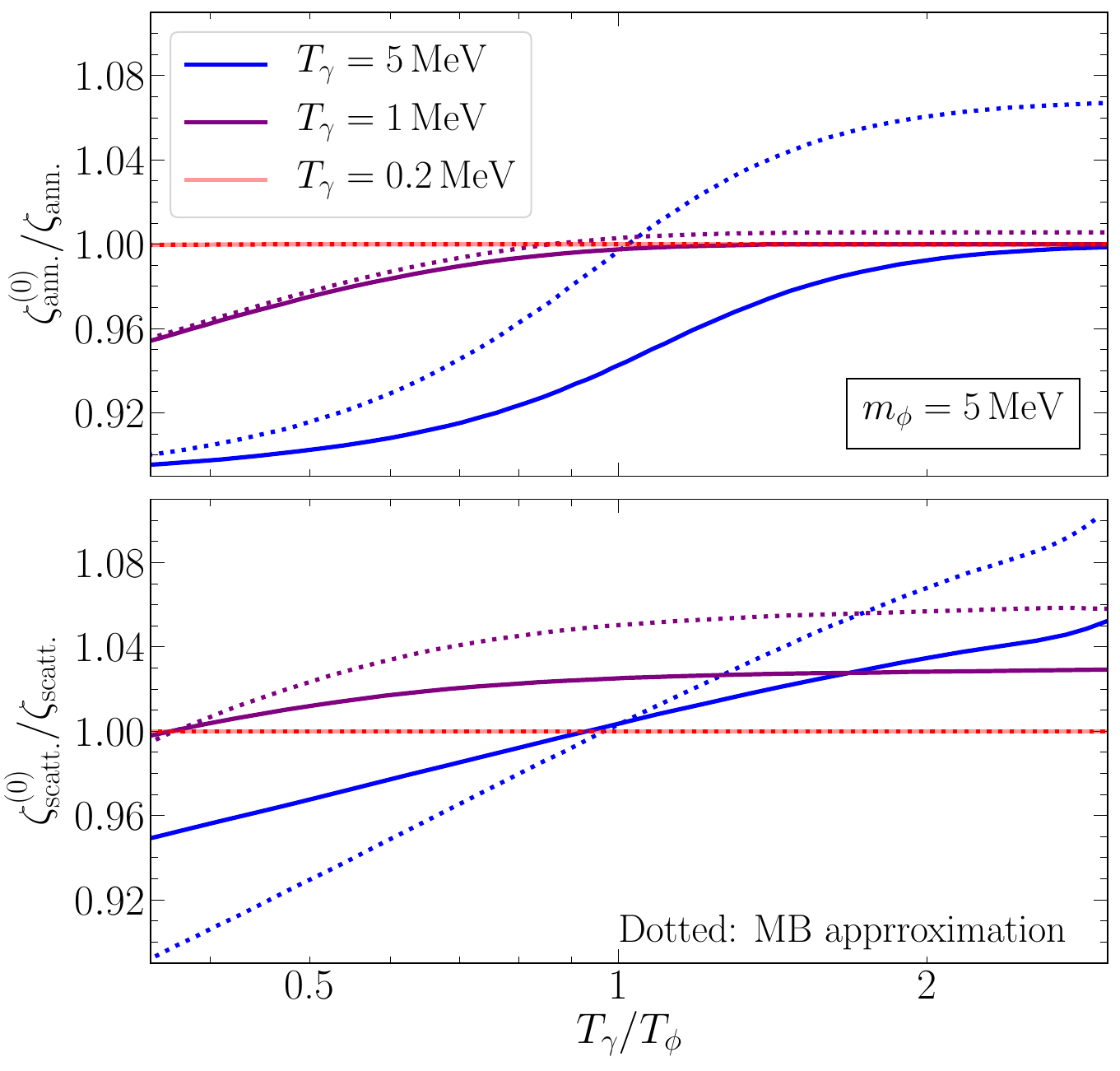}
\caption{The ratio of  leading-order to exact collision terms as a function of the temperature ratio  $T_\gamma/T_\phi$ for  annihilation (upper panel) and elastic scattering (lower panel) is shown as solid lines; $m_\phi =5$\,MeV is chosen for both panels. The classical limit adopting the Maxwell-Boltzmann (MB) distribution functions is also shown by the dotted lines for comparison. The various colors depict  different epochs of the Universe, characterized by $T_\gamma$. In the  non-relativistic limit  $T_{\phi,\gamma} \ll m_\phi$, both approximations converge to the exact results. We exclude the nominal singular point at $T_\phi = T_\gamma$, where all rates vanish.}
\label{fig:uncertain}
\end{center}
\end{figure}

Before going further to present the numerical results, we illustrate and comment on  the uncertainties introduced by our approximations. 

One uncertainty comes from our approximated interaction rates, which neglect final state statistics. The differences from the exact rates are shown in Fig.~\ref{fig:uncertain} for the energy transfer between the electron and dark scalar for the choice  $m_\phi = 5$\,MeV. The solid lines show the interaction rates by taking $1\pm f_{3,4} \simeq 1$, normalized to the exact values at different photon temperatures (illustrated with different colors), suggesting percent-level uncertainties after neutrino decoupling for $m_\phi \ge 5$\,MeV. Our method improves the precision of interaction rates, especially at $T_\phi \le m_\phi/3$,  with respect to the classical approximation (Maxwell-Boltzmann distributions, dotted lines) that is typically adopted in the literature. In the deep non-relativistic regimes where $T_{\gamma, \phi} \ll m_{\phi}$, all lines converge, as expected.  Similar results apply to neutrino-$\phi$ interactions.  

Another source of uncertainty is introduced by parametrizing each momentum distribution by two variables, $T_i$ and $\mu_i$, only.  This is of course perfectly justified for the electron sector. For the neutrino sector, the non-thermal contribution is below the percent-level~\cite{Mangano:2005cc, deSalas:2016ztq}, and it is expected to be even smaller with flavor-blind dark sector-neutrino interactions. When we present our numerical results below, we will demonstrate that we recover the state-of-the-art prediction for $N_{\rm eff}$. Finally, a  description in terms of temperature and chemical potential in the dark sector is expected to give accurate results for the problem at hand as well. The reason is that  kinetic equilibrium through self-scattering is typically maintained until after freeze-out. The interactions of the dark sector with itself and with SM particles require specification of the particle physics model, of which we now present an example and for which these conditions are satisfied.

\section{Representative Particle Model}
\label{sec:particle_model}

For the quantitative exploration we shall consider a complex scalar DM candidate $\phi$ with a $Z'$ mediator~\cite{Boehm:2003hm,Boehm:2020wbt} as the representative particle physics model. The setup has been studied in detail as a potential explanation of both the INTEGRAL $511\,{\rm keV}$ line and muonic $g-2$ anomaly~\cite{Boehm:2003bt}, as well as for its signatures in intensity-frontier experiments~\cite{Boehm:2020wbt}.
The interactions between $Z'$, $\phi$ and SM leptons $l=e,\nu_e, \dots$ are given by 
\begin{align}
 \mathcal{L}_{Z'}^{\rm int} &= g_\phi^2 Z'^\mu Z'_\mu \phi^* \phi - i g_\phi Z'^\mu (\phi^*\overset{\leftrightarrow}{\partial}_\mu \phi)   -g_l Z'^\mu \bar{l} \gamma_\mu l\,.
\end{align}
For concreteness, we consider flavor-blind couplings $g_l$ and assume for the $Z'$ mass $m_{Z'} \gtrsim 1\,{\rm GeV}$ such that $Z'$ remains off-shell in the scattering/annihilation processes and has negligible population during DM freeze-out. In further consequence, the interactions between $\phi$ and SM leptons can be treated as by an effective operator with UV scale $\Lambda_{Z'} = m_{Z'}/\sqrt{g_\phi g_l}$ and it is in this quantity how we shall present our results.  Note that in this specific model, DM freeze-out is dominated by $p$-wave annihilation.  Furthermore, we choose  $g_\phi \gg g_l$ so that $Z'$-mediated $e-\nu$ interactions can be neglected. An exploration of non-flavor-blind couplings and variable branching ratios will be presented in a dedicated work~\cite{companion}. Finally, also note that there is no tree level interaction between $\phi/Z'$ and the photon. An extension to scenarios where the coupling to the EM sector is (also) through $\gamma$ is straightforward.

\section{Solution of three-sector system}
\label{sec:numsolution}

Having established the full formulation of the problem,  we are now in a position to numerically solve the set of Boltzmann equations for the three-sector system. We do so by solving for the evolution of the five variables $T_\gamma$, $T_\nu$, $T_\phi$, $\tmu_\nu$ and $\tmu_\phi$, as functions of time, or, equivalently, the evolution of $\rho_{\rm EM}$, $\rho_\nu$, $\rho_\phi$, $n_\nu$ and $n_\phi$.%
\footnote{The correspondence between  ($\rho_i$, $n_i$) and  ($T_i$, $\mu_i$) is given in App.~\ref{sec:conversion}.}
For this we calculate and tabulate the values of $\gamma_{ij} (T_i, T_j)$ and  $\zeta_{ij} (T_i, T_j)$ for each process on a grid of temperatures $(T_i,\, T_j)$.
The tabulations are then used to compute the evolution of number- and energy-densities, $n_i$ and $\rho_i$, of the three sectors, respectively.\footnote{For the SM processes, we have applied finite temperature QED corrections~\cite{Heckler:1994tv,Fornengo:1997wa}  tabulated in \texttt{nudec\textunderscore BSM}~\cite{Escudero:2018mvt}.} As mentioned above, the advantage of such tabulation is that it only needs to be done once for obtaining the evolution of $\delta n_i/\delta t$ and $\delta \rho_i/\delta t$ once $m_\phi$ is fixed; the actual rates are found by  rescaling with the chemical potential of each species, $\tmu_i$, as well as the UV scale of our benchmark model, $\Lambda_{Z'}$. 

\subsection{SM-only solution (\boldmath $N_{\rm eff}^{\rm SM}$ prediction)}

 \begin{figure}[!t]
\includegraphics[width=\columnwidth]{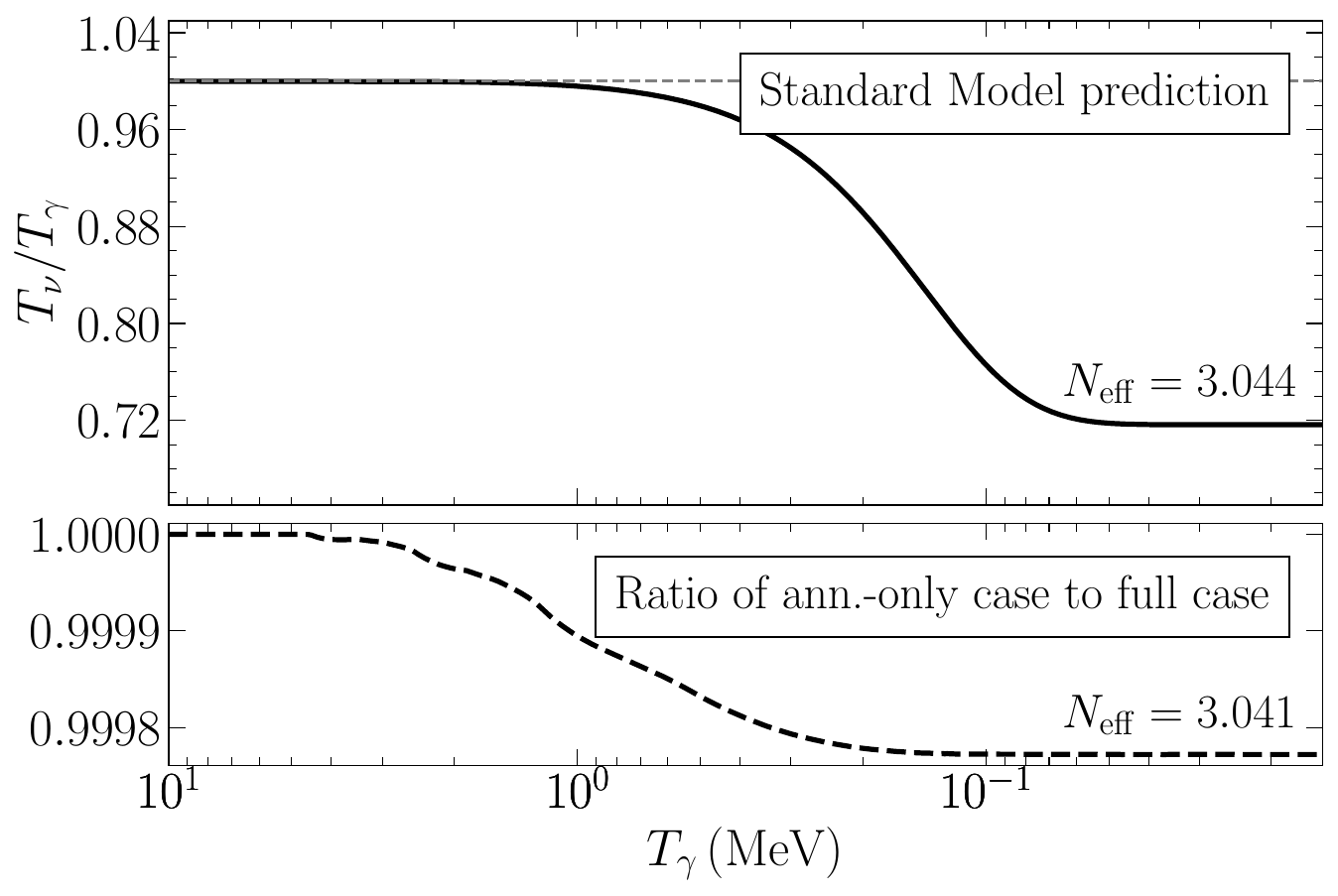}~~
\caption{
The SM evolution of temperature ratio $T_\nu/T_\gamma$ (without dark sector interactions). The calculation of interaction rate is based on our formalism, with NLO-QED corrections adopted from the code developed in~\cite{EscuderoAbenza:2020cmq}. We obtain the final value of $N_{\rm eff} = 3.044$ (solid line in upper panel) in concordance with the recently-reported value $3.0440\pm 0.0002$ in \cite{Bennett:2020zkv}.
In the lower panel, we show $T_\nu/ T_\gamma$ ratio of without $\nu$-$e$ scattering case to the full case; $\nu$-$e$ scattering affects the $N_{\rm eff}$ prediction only in the third digit.}
\label{fig:SMonly_Tevo}
\end{figure}

Before studying the interplay of the dark sector with the SM quantities, we  compare our numerical prediction of the standard cosmology, \textit{i.e.}, turning off the dark sector, with those in the literature that solve for the general neutrino momentum distributions. The final result is reported in terms of the  parameter $N_{\rm eff}^{\rm SM}$. It counts the relativistic SM neutrino  degrees of freedom, that besides the two photon polarizations constitute the standard radiation content after BBN and prior to recombination. As such, it  measures to which degree the neutrino-to-photon temperature ratio deviates from the value derived from SM entropy conservation, $(4/11)^{1/3} \simeq 0.7138$.

Figure~\ref{fig:SMonly_Tevo} shows our result for the evolution of $T_\nu/T_\gamma$ as a function of photon temperature. As can be seen, neutrinos decouple from the EM sector at $T_{\gamma} \sim 2$\,MeV. 
For $T_\gamma\lesssim 40$~keV,
electron annihilation completes and the ratio $T_\nu/T_\gamma $ freezes out. As final value we obtain $N_{\rm eff}^{\rm SM} = 3.044$ in concordance with the values reported in the recent literature, $3.043-3.046$~\cite{deSalas:2016ztq, Bennett:2019ewm, EscuderoAbenza:2020cmq, Froustey:2020mcq, Akita:2020szl,  Bennett:2020zkv}; neutrino oscillations induce a correction to $N_{\rm eff}$ at the  $0.001$ level only~\cite{Mangano:2005cc}.  
Since $\nu$-$e$ scattering maintains the kinetic coupling for a short period even after chemical decoupling, neglecting the scattering leads to smaller $T_\nu/T_\gamma$ and $N_{\rm eff}$, as shown in the lower panel of Figure~\ref{fig:SMonly_Tevo}. A similar numerical difference induced by scattering processes has  been obtained in the literature, see \textit{e.g.}~\cite{Grohs:2015tfy}. 

Having recovered the standard prediction gives credence to our approach and shows that any introduced errors through our approximate treatment are indeed well under control.

\begin{figure*}[t]
\begin{center}
\includegraphics[width=0.48\textwidth]{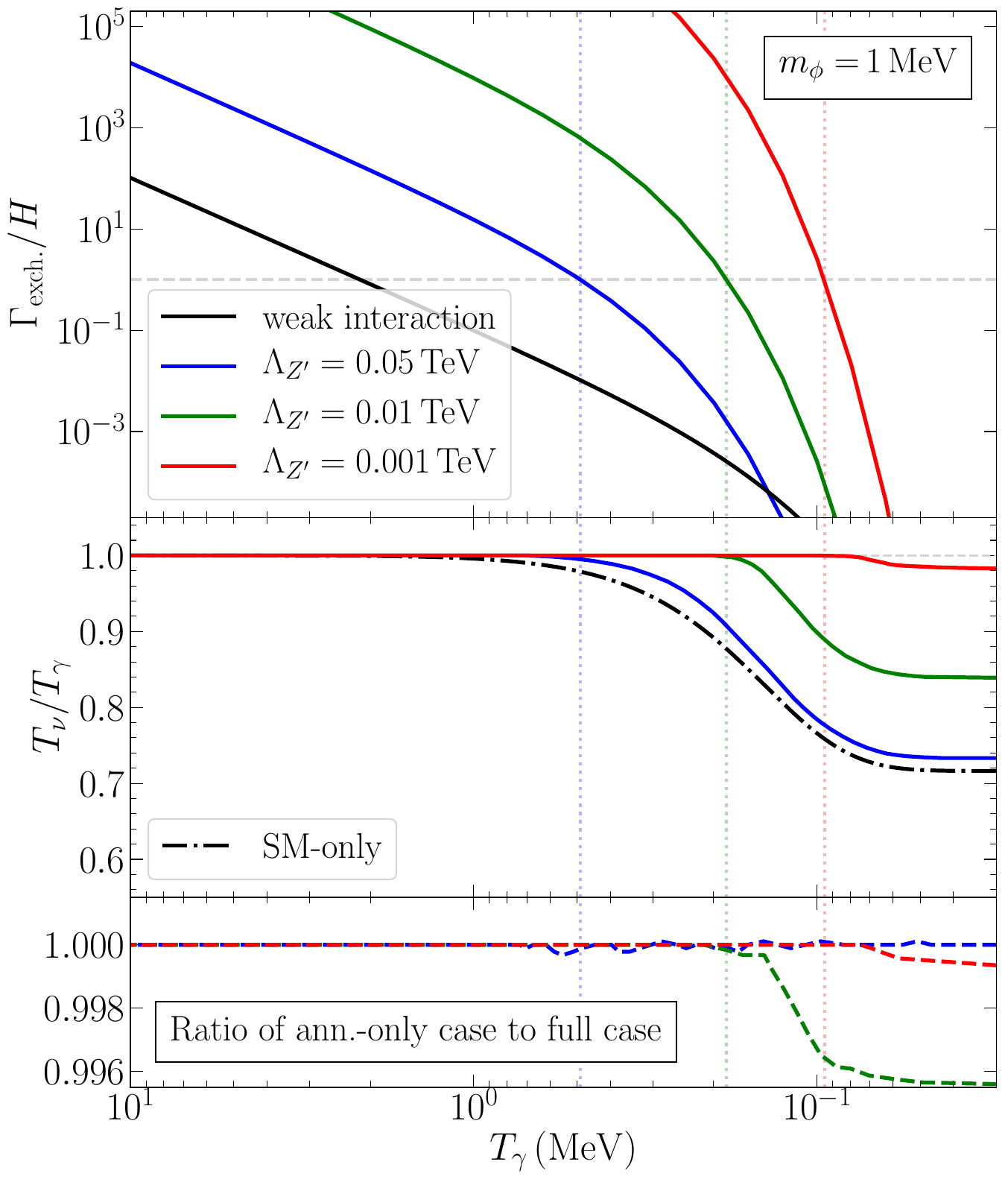}\hfill
\includegraphics[width=0.48\textwidth]{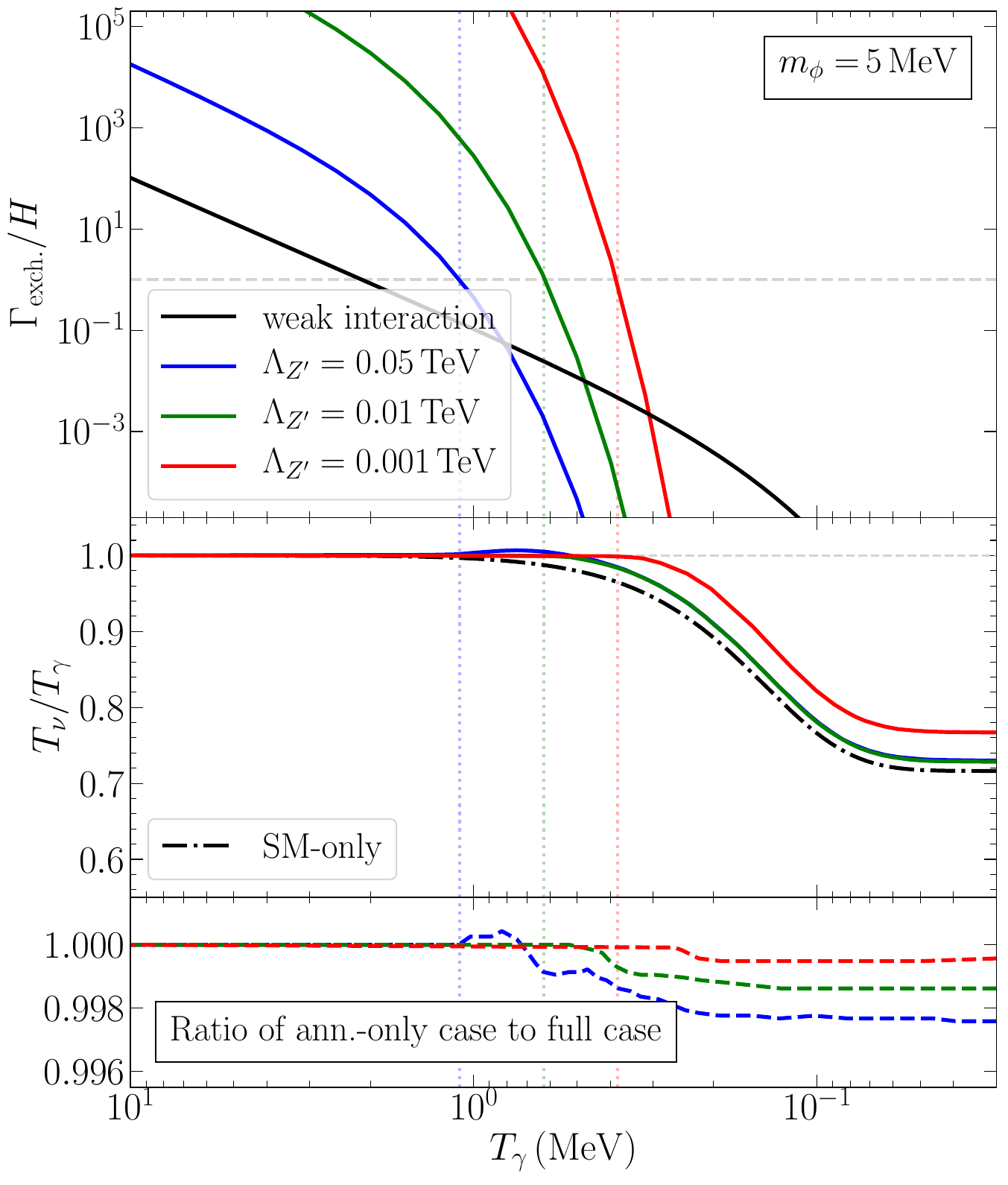}
\end{center}
\caption{The ratio of the $\phi$-induced energy-transfer rate to the Hubble rate, $\Gamma_{\rm exch.}/H$ (top panels), the evolution of $T_\nu/T_\gamma$ (middle panels), and the latter's difference when scattering is excluded (bottom panels) as a function of photon temperature $T_\gamma$. The left (right) panel shows the result for $m_\phi= 1 (5)\,{\rm MeV}$ and various choices of $\Lambda_{Z'}$ as labeled. 
For comparison we also show the ratio of  $\Gamma_{\rm weak}/H$  by the black solid lines. 
The vertical dotted lines depict the  $T_\gamma$ points where $\Gamma_{\rm exch.}/H = 1$.
Decreasing $\Lambda_{Z'}$ keeps neutrino and EM sectors longer in equilibrium.
}
\label{fig:DM_Tnuevo}
\end{figure*}
\begin{figure*}[t]
\begin{center}
\includegraphics[width=0.48\textwidth]{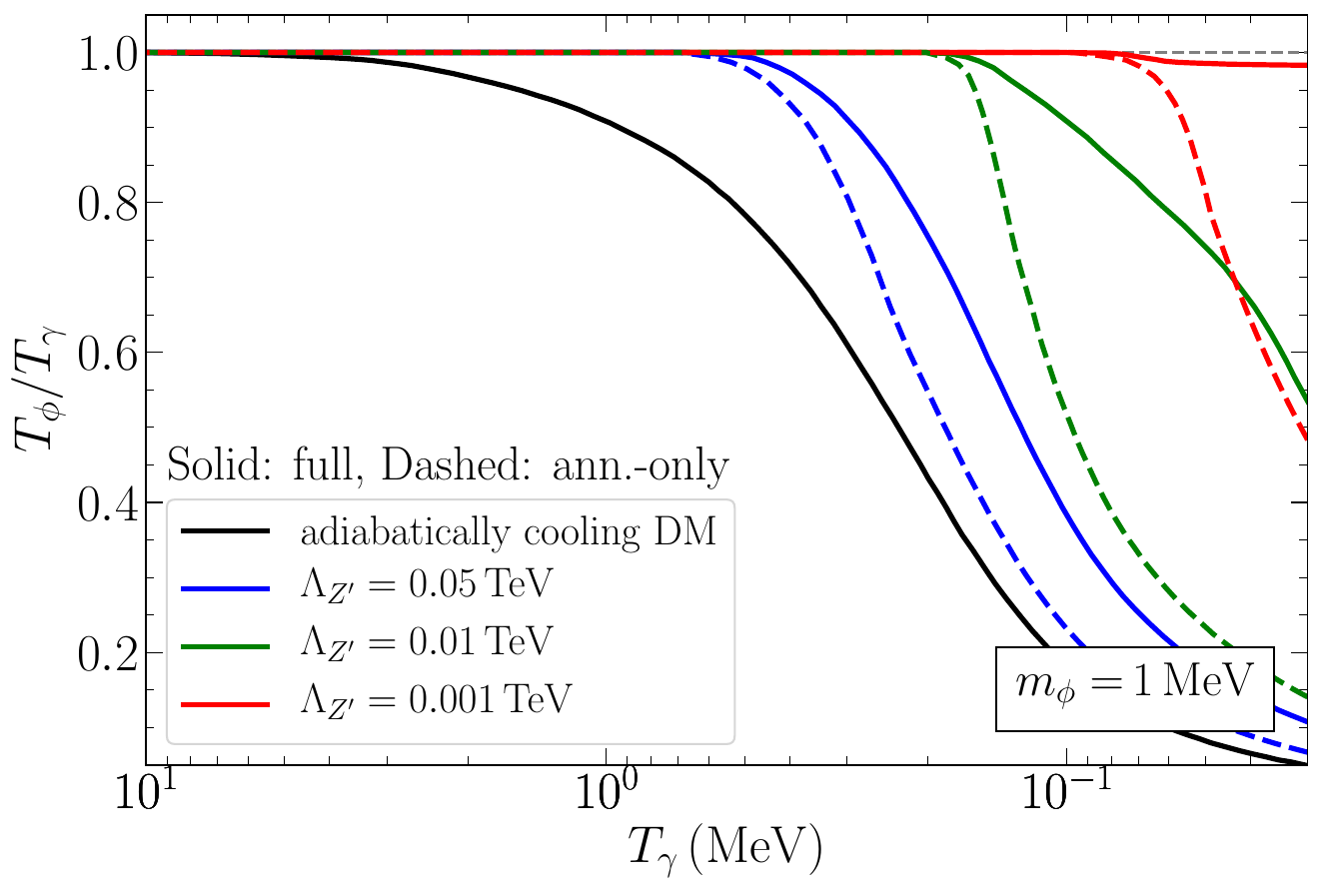}
\includegraphics[width=0.48\textwidth]{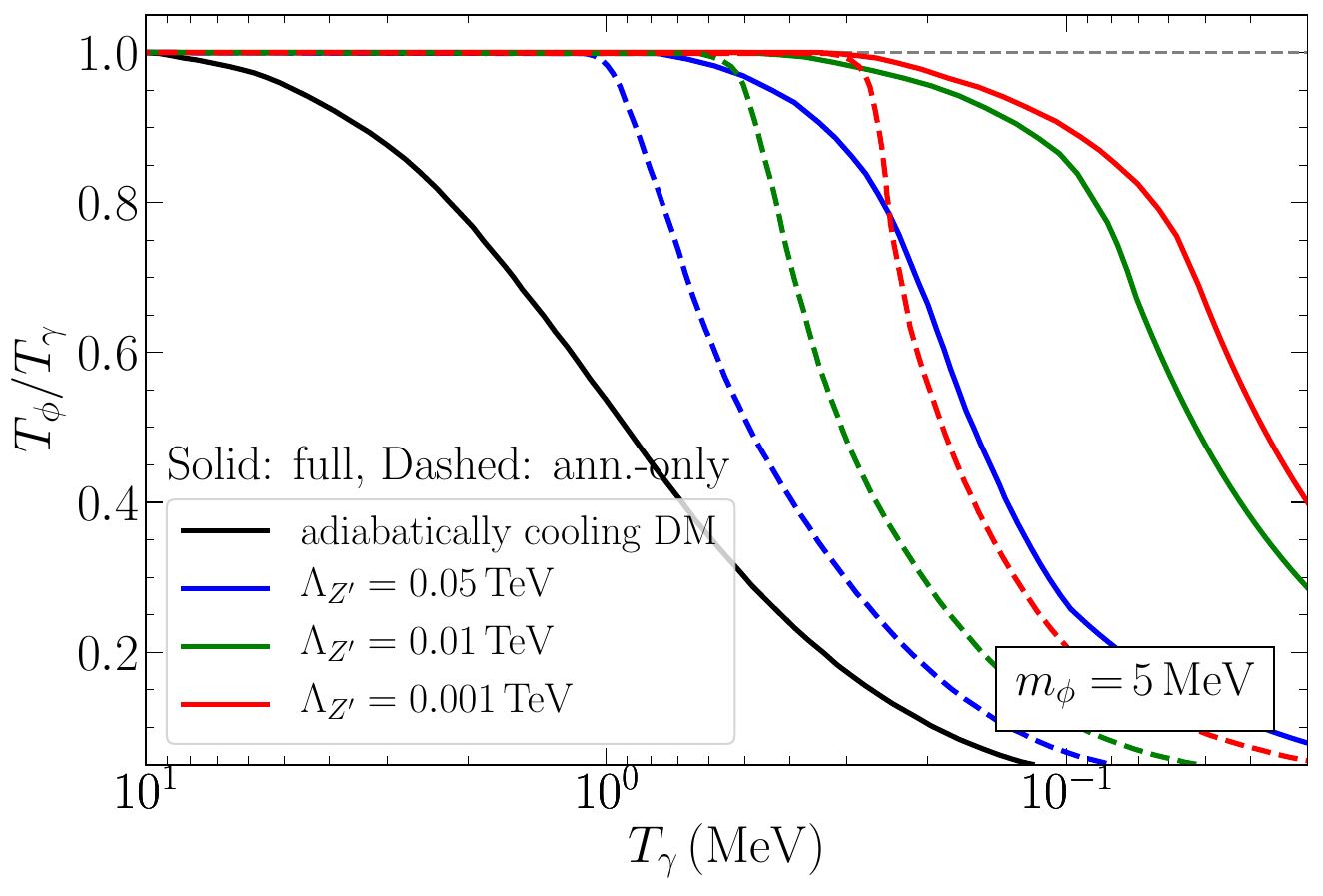}
\end{center}
\caption{The evolution of $T_\phi/T_\gamma$ for $m_\phi = 1\,{\rm MeV}$ (left panel) and $5\,{\rm MeV}$ (right panel). The solid lines show the case including $\phi$ annihilation and scattering, while the dashed lines only takes into account  annihilation.
We also show temperature evolution of an adiabatically cooling DM species with vanishing chemical potential and that is assumed to kinetically decouple from photons at $T_\gamma = 10\,{\MeV}$ (black solid line).
}
\label{fig:DM_Tphievo}
\end{figure*}
\begin{figure*}[!t]
\begin{center}
\includegraphics[width=0.48\textwidth]{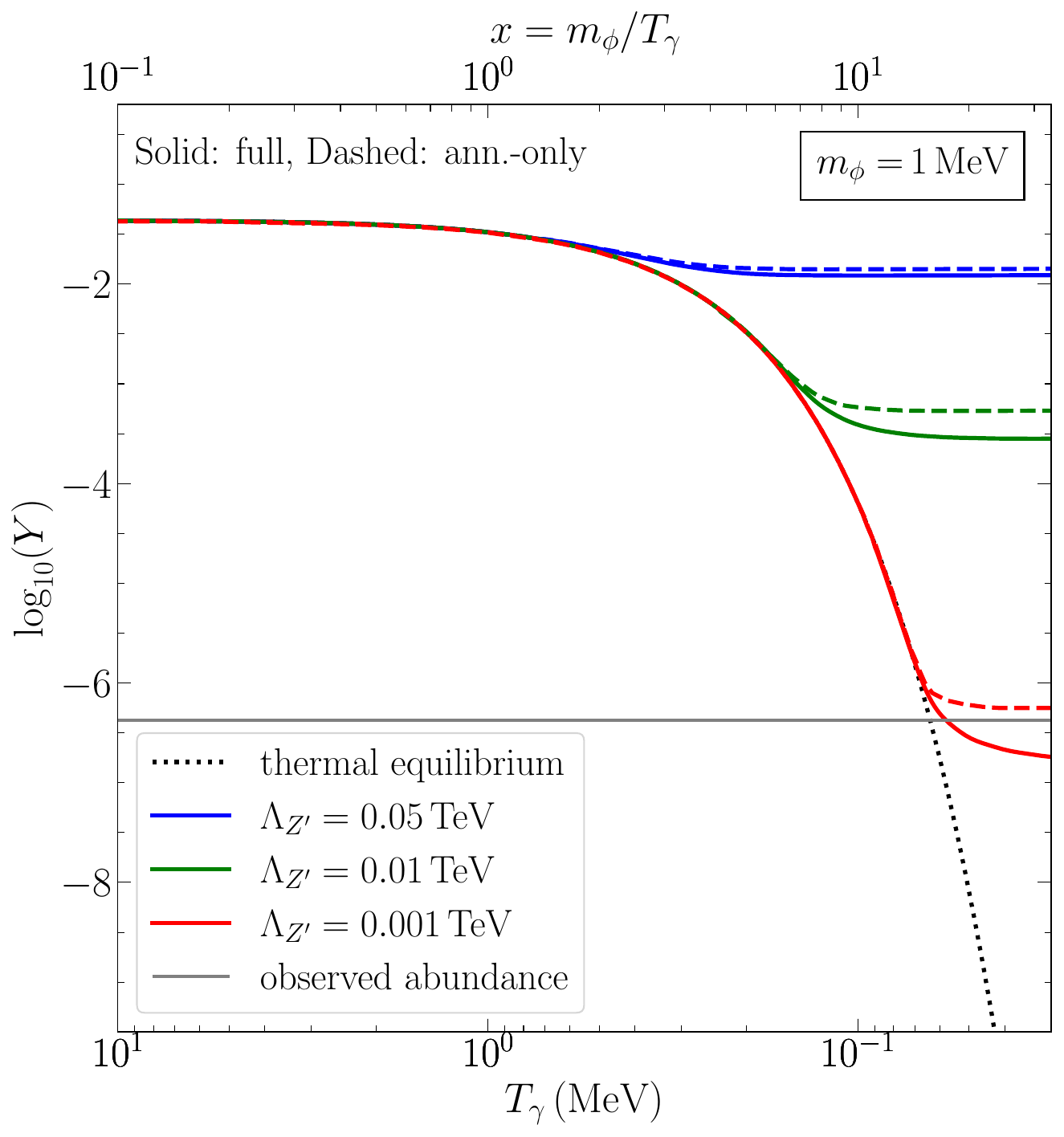}
\includegraphics[width=0.48\textwidth]{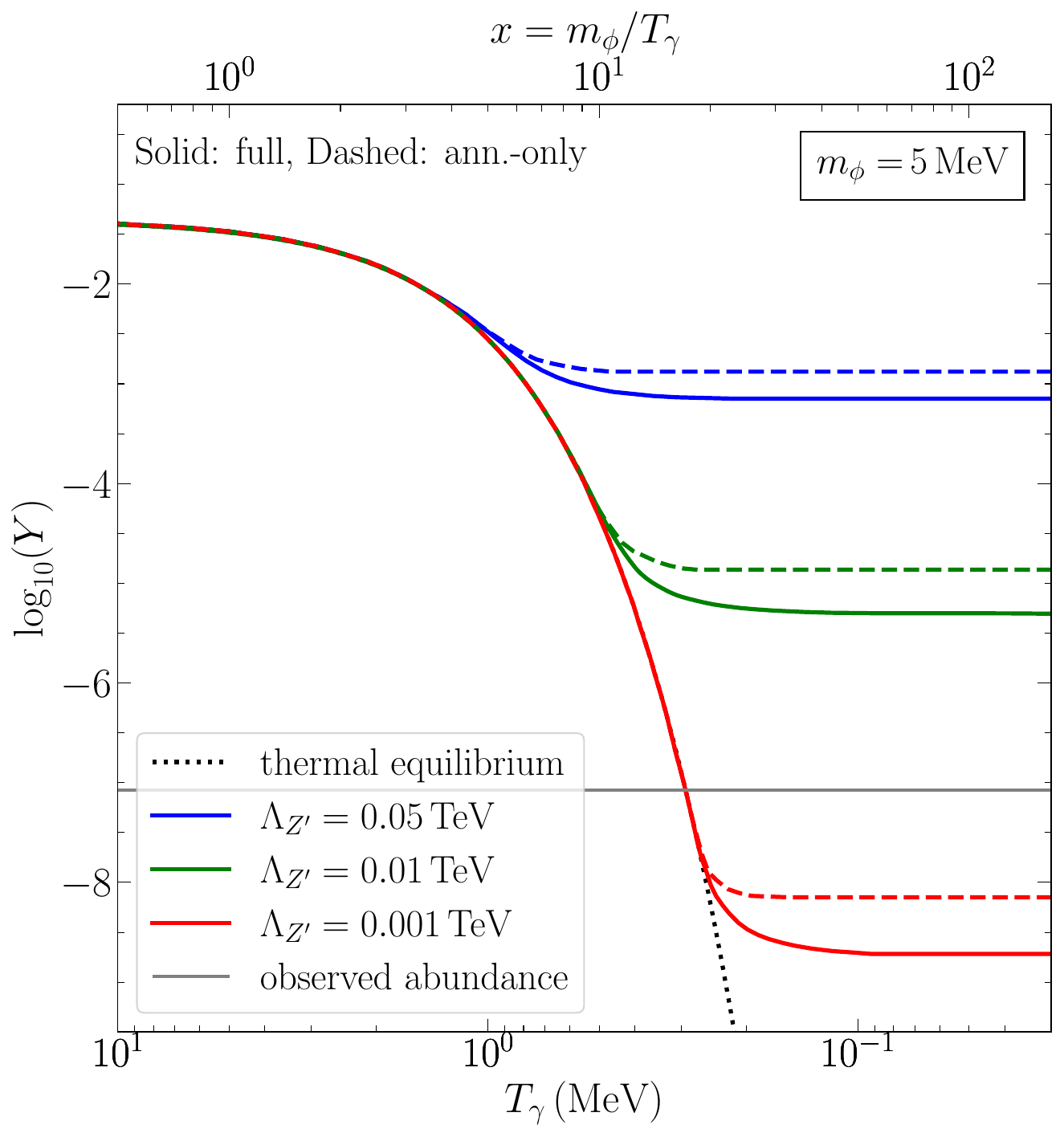}
\end{center}
\caption{
Evolution of the DM abundance $Y$ as a function of $T_\gamma$ for $m_\phi = 1\,{\rm MeV}$ (left panel) and $5\,{\rm MeV}$ (right panel). 
The observed DM abundance is shown as gray solid lines for comparison; see Eq.~\eqref{eq:correct_relic_abundance}. The dashed lines neglect elastic scattering from the treatment. This leads to a lower dark sector temperature $T_\phi$ and less efficient $p$-wave annihilation, resulting in higher DM yields.
}
\label{fig:DM_abundance}
\end{figure*}
\subsection{Neutrino  temperature evolution}
\label{sec:evo_temp}

 We now proceed including $\phi$ and study the joint evolution with the EM and $\nu$ sectors. The initial temperature  is set to $T_\gamma = 10\,$MeV and we assume  $ \mu_\nu = \mu_\phi = 0$ as thermal initial conditions.
At this early stage, if the normalized energy-transfer rate induced by $\phi$ defined in Eq.~\eqref{rates} is much larger than the Hubble expansion rate,  $\Gamma_{\rm exch.}  \gg H$ --- equivalent to saying that, within a Hubble time, at least an amount of $\rho_\gamma$ or  $\rho_\nu$ is exchanged between the $\phi$-sector with the respective EM and $\nu$ sectors --- we can safely take all three sectors to be in thermal equilibrium and all particle species share the same temperature.\footnote{In this case, even if weak interactions freeze out, DM-SM interactions can keep EM and neutrino sectors in thermal equilibrium~\cite{Depta:2019lbe}.} 
As stated above, the situation is different from WIMPs,  where the energy exchange with the EM and $\nu$-sectors can be neglected at MeV temperatures.
Here, all three sectors evolve independently  once both $H \gg \Gamma_{\rm exch.}, \Gamma_{\rm weak}$ are fulfilled.

The top panel of Fig.~\ref{fig:DM_Tnuevo}  shows the ratios $\Gamma_{\rm exch.}/H$ and $\Gamma_{\rm weak}/H$  for the choices $\Lambda_{Z'} = 0.05,\ 0.01, \ 0.001$~TeV.
The left and right panels are for $m_\phi = 1$~MeV and  $m_\phi = 5$~MeV, respectively. 
We observe that $\Gamma_{\rm exch.}> \Gamma_{\rm weak}$ so that  $\Gamma_{\rm exch.}$ controls the decoupling of photon and neutrino temperatures. 
For example, for $m_\phi = 1\,$MeV and $\Lambda_{Z'} = 0.05$\,TeV (blue lines) we find $\Gamma_{\rm exch.}/H=1$ at $T_\gamma \simeq 0.5$\,MeV. The middle panel of Fig.~\ref{fig:DM_Tnuevo} shows the corresponding evolution of $T_\nu/T_\gamma$ and around $T_\gamma = 0.5~$MeV neutrino and photon temperatures can be seen to depart from each other for this parameter set. 
The figures also show that stronger SM-DM interactions, i.e., smaller values of $\Lambda_{Z'}$ (green/red lines), allow for prolonged dark sector-mediated energy exchange. In consequence, $T_\nu$ deviates from $T_\gamma$ {\it later} in comparison to a standard cosmological history. 
Later decoupling is also observed for decreasing DM mass. 
This is because, for heavier DM, its number density becomes Boltzmann suppressed earlier, compensating for any rise in the  DM annihilation cross section proportional to $m_\phi^2/\Lambda_{Z'}^4$.

\begin{figure*} 
\includegraphics[width=0.49\textwidth]{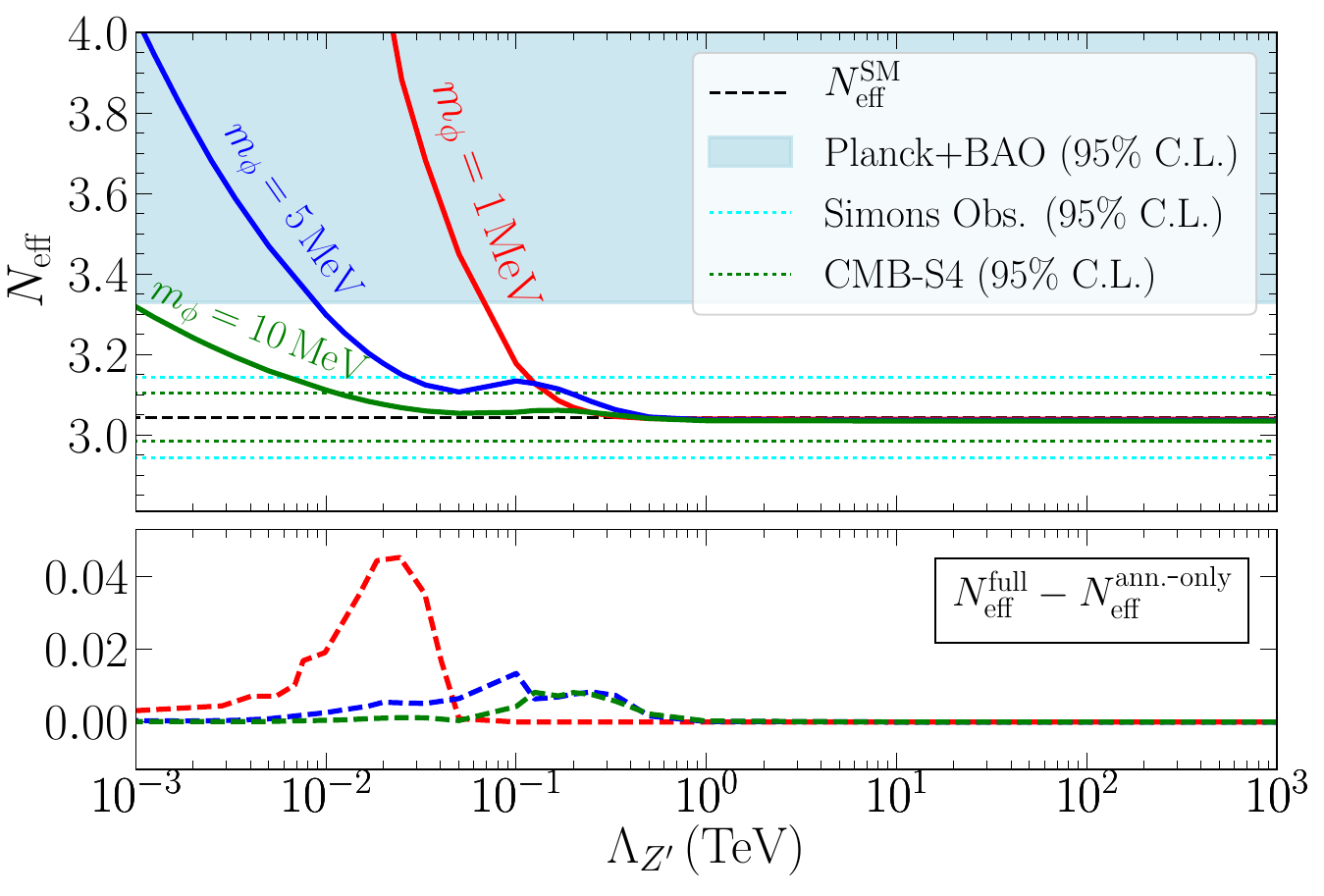}
\includegraphics[width=0.49\textwidth]{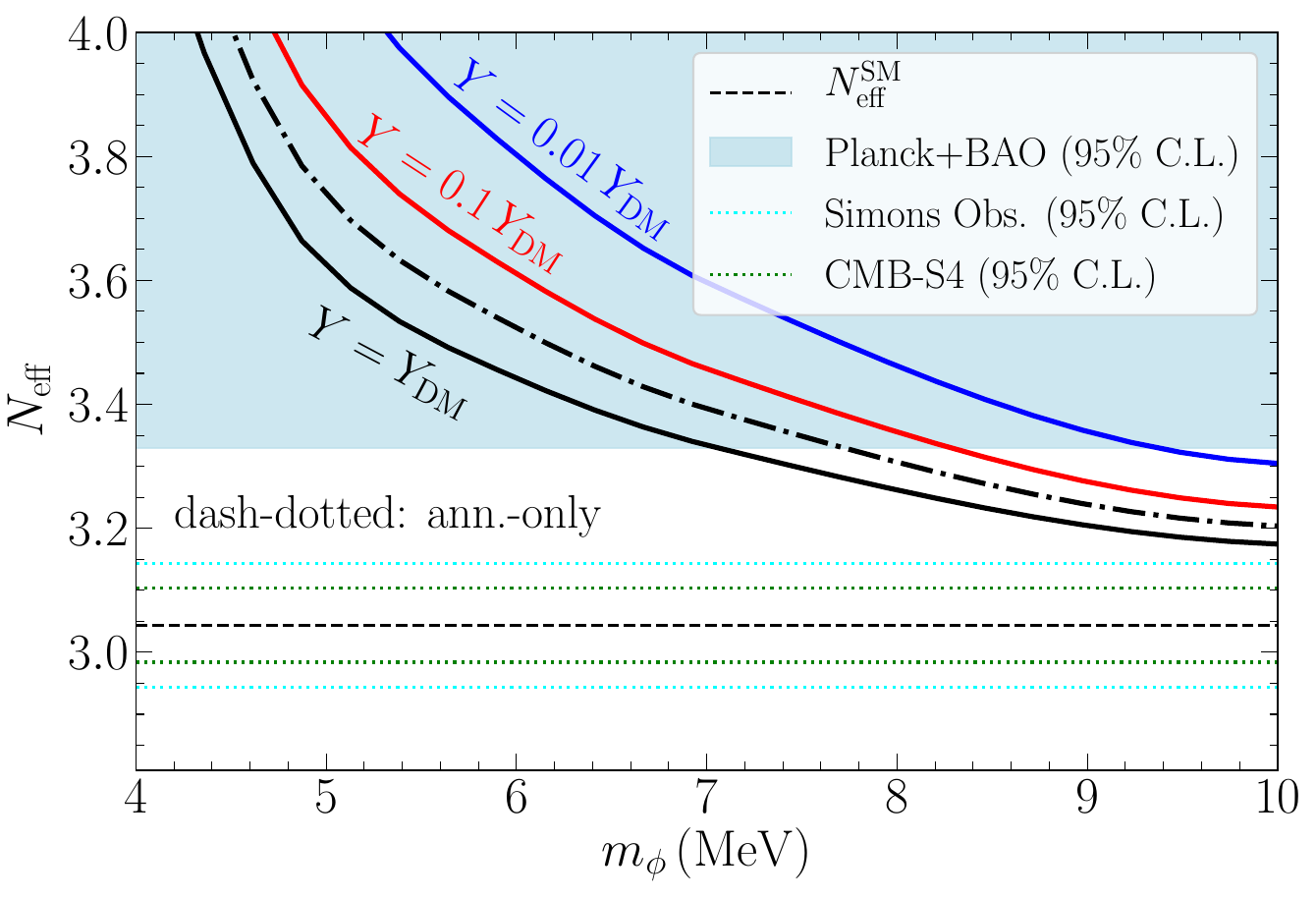}
\caption{\textit{Left panel}:
The top panel shows the prediction of $N_{\rm eff}$ as a function of $\Lambda_{Z'}$ for varying DM masses as labeled together with the present CMB constraint (shaded) and future sensitivity (dotted lines). The bottom panes shows the (minute) differences in $N_{\rm eff}$ from the full result, when scattering processes are turned off.
\textit{Right panel}: $N_{\rm eff}$ prediction as a function of $m_\phi$ imposing a fractional relic density as labeled.
The current cosmological limit excludes a thermal relic (black solid line) for $m_\phi \lesssim 7\,{\rm MeV}$ for the chosen benchmark particle model, compared to $m_\phi \lesssim 7.6\,{\rm MeV}$ if one neglects scatterings (black dash-dotted line). If $\phi$ only constitutes $10\%$ ($1\%$) of DM, one finds that $m_\phi \lesssim 8.3\, (9.4)\,{\rm MeV}$ is excluded. 
}
\label{fig:Neff}
\end{figure*}

After the $ \phi$-induced energy exchange processes between the neutrino and EM sectors decouple at $\Gamma_{\rm exch.}/H \sim 1$, the residual annihilation of $\phi$ particles continues to heat both sectors separately. In the considered flavor-blind model, this happens with  a slight preference for neutrinos. Such residual annihilation can be important in increasing $N_{\rm eff}$ only if $\rho_\phi/\rho_\nu$ is still sizeable at $\Gamma_{\rm exch.}/H \sim 1$ and DM  freezes out non-relativistically subsequently.\footnote{That is, this effect is only visible for a limited range of DM mass and $\Lambda_{Z'}$, and explains, in the left upper panel of Fig.~\ref{fig:Neff} later, the bump at $\Lambda_{Z'}\sim 0.2$\,TeV for $M_\phi = 5$\,MeV (the blue line).} 
It is best observed for $m_\phi =5$~MeV where even $T_\nu/T_\gamma \ge 1$ becomes possible for an intermittent period of time. For $m_\phi = 1$~MeV, such heating is very mild, especially when the DM freeze-out happens almost relativistically, corresponding to $\Lambda_{Z'} \gtrsim   0.05\,$TeV. 
The overall effect is that $T_\nu/T_\gamma$ decreases less slowly than in the SM-only case. 
Finally, the $\phi$ annihilation rate becomes Boltzmann suppressed and $e^{\pm}$ annihilation takes over, heating primarily photons.  This leads to the usual decrease of $T_\nu/T_\gamma$ and the ratio freezes out after both $e^{\pm}$ and $\phi\phi^*$ annihilation processes become negligible. 

In the bottom panel of Fig.~\ref{fig:DM_Tnuevo} we explore the effect of the elastic scattering contribution on the neutrino temperature by plotting the ratio of $T_\nu/T_\gamma$ without scattering to the one obtained including scattering (full result). As can be seen, for symmetric thermal DM, the effect on the neutrino temperature is below percent-level although it would enter in a precision determination of $N_{\rm eff}$. In other words, the energy transfer efficiency of $\phi$-annihilation typically dominates over that of $\phi-$SM scattering in the evolution of $T_\nu/T_\gamma$. The situation is markedly different for $T_\phi$. There, elastic scattering enters in an important way, and we shall discuss the evolution of $T_\phi/T_\gamma$ next.

\subsection{DM temperature and abundance evolution}
\label{sec:evo_tempDM}

Turning to the evolution of $T_\phi$, we note that  DM-SM scattering may keep DM in kinetic equilibrium with its annihilation products after DM freeze-out. However, since at that point $T_\nu\neq T_\gamma$, the evolution of $T_\phi$ becomes subtle and $T_\phi$ will in general lie {\it between} $T_\nu$ and $T_\gamma$ before DM  decouples kinetically.
Figure~\ref{fig:DM_Tphievo} shows the evolution of $T_\phi/T_\gamma$,  where the solid lines are the full result and dashed lines ignore the elastic scattering contribution. As can be seen, scattering leads to a slower decline in $T_\phi$ since $\Gamma_{{\rm scatt.},i}> H$.
The $T_\phi$ evolution enters the abundance determination as will be shown below.

We are now in a position to solve for the DM abundance which we report in terms of the yield variable $Y\equiv (n_{\phi}+n_{\phi^*})/s$, including both particle and anti-particle number densities,  with $s$ being the entropy density. When $\phi,\ \phi^*$ makes all the DM, it must match the observationally inferred abundance~\cite{Planck:2018vyg}, 
\begin{align}
\label{eq:correct_relic_abundance}
Y_{\rm DM} &= \dfrac{\Omega_{\rm DM} \rho_{\rm crit} }{m_\phi s_0} =4.2 \times 10^{-7} \left(\dfrac{\rm MeV}{m_\phi}\right) \,.
\end{align}
Here, $\Omega_{\rm DM}$, $\rho_{\rm crit}$ and $s_0$ are the DM density parameter, the critical density of the Universe and the current entropy density, respectively.
While $\mu_\nu$ is always very small as neutrinos are strictly relativistic in the early Universe, the evolution of $\mu_\phi$ is non-trivial when DM  becomes non-relativistic. In this final, non-relativistic period, the DM number density gradually freezes out when its chemical potential converges towards $m_\phi$ and $Y$ becomes a constant. 

After weak interactions decouple, the  dark sector evolution can be separated into four periods, the first three of which were depicted earlier in Fig.~\ref{fig:scheme}: $i)$ At early times, when the $\phi$-induced energy-exchange rate  is larger than the Hubble rate, $\Gamma_{\rm exch.} >H$, and all three sectors share the same temperature. Since $\rho_\gamma \ge \rho_\phi \simeq n_\phi \langle \delta E \rangle $, the condition $\Gamma_{\rm ann.} > H$ is automatically satisfied. That is, as expected, the evolving $\phi$  abundance follows the chemical equilibrium value, $Y_\phi^{\rm eq} \equiv n_\phi^{\rm eq}/s$, characterized by the common temperature. 
 $ii)$  Once $\Gamma_{\rm exch.}$ drops below $H$ but $\Gamma_{\rm ann.} >  H$ remains true, the  $\phi$ abundance continues to follow $Y_\phi^{\rm eq}$.  Typically, $\phi$ scatters efficiently with both EM and neutrino sectors in this period, $\Gamma_{{\rm scatt.},i} >H$ with $i={\rm EM},\nu$, and  $T_\phi$ lies between $T_\gamma$ and $T_\nu$ with its detailed value depending on the relative interaction strength; note that once electrons become Boltzmann suppressed, one encounters a steep decline in $\Gamma_{{\rm scatt.,EM}}$ unless a direct interaction with photons in invoked.\footnote{For instance, in the case of $m_\phi =1\,$MeV with $\Lambda_{Z'} = 0.001$\,TeV (red lines in right panels),  $\Gamma_{{\rm exch.,EM}} \sim H$ occurs  at $T_\gamma \sim 0.1$\,MeV, below which $T_\phi$ simply follows $T_\nu$, as there are too few $e^\pm$ left.} 
 $iii)$ Later, $\Gamma_{\rm ann.} < H$ becomes fulfilled, and DM chemically decouples from the SM thermal bath.
The decoupling of DM is not instantaneous but with a short transition period, during which $\mu_\phi$ increases from a negligible value to approximately $m_\phi$, resulting in the scaling $n_\phi \propto a^{-3}$ with $a$ being the scale factor in comparison to exponential suppression of $n_\phi^{\rm eq}$.  
$iv)$ Finally,  $ \Gamma_{{\rm scatt.},i}<H $ and DM kinetically decouples from sector~$i = {\rm EM}$ and/or $\nu$.\footnote{In the case of (semi-)relativistic freeze-out,  thermal and kinetic decoupling would happen around the same time.} 
After kinetic decoupling, DM  adiabatically cools, $T_\phi \propto a^{-2}$, or, equivalently, $T_\phi/T_\gamma \propto T_\gamma$. 

The evolution of the DM abundance as a function of $T_\gamma$ is shown in Fig.~\ref{fig:DM_abundance}. Weaker interactions lead to earlier decoupling and higher abundances. The solid lines show the full result and the dashed lines show the result with the elastic scattering contribution turned off. The annihilation cross sections required for obtaining the correct value of $\Omega_\text{DM}$ can thus differ by a factor of a few, as DM-SM elastic scattering extends the thermalized-to-decoupled transition. This is particularly true for  $p$-wave annihilation as its efficiency depends on the dark sector temperature in an elevated way.  It demonstrates the importance to keep track of the individual temperatures. To our knowledge, this has not been demonstrated previously as most (semi-)analytic or even entirely numeric calculations of relic density assume that DM remains kinetically coupled to the thermal bath of all its annihilation products (which are all assumed to share one temperature).

\section{The CMB \boldmath$N_{\rm eff}$ constraint}
\label{sec:deltaNeff}

A central merit of our introduced methodology is that it opens the door for a precision prediction of $N_{\rm eff}$ for light DM annihilation into the EM and $\nu$ sectors, {a priori} with {\it arbitrary} branching ratios. So far, this has not been possible with the available treatments in the literature. The departure from the SM-only prediction is parameterized as~\cite{Planck:2018vyg}
\begin{align}
  \Delta N_{\rm eff} \equiv N_{\rm eff} - N_{\rm eff}^{\rm SM}\,,
\end{align}
and we adopt $N_{\rm eff}^{\rm SM}= 3.044$  consistent with the result above. At 95\% C.L.~the
combination of Planck and BAO measurements yields $ 2.66 \leq N_{\rm eff} \leq 3.33$~\cite{Planck:2018vyg}. This number is expected to improve by the future Simons Observatory to $|\Delta N_{\rm eff}|\lesssim 0.1$~\cite{SimonsObservatory:2018koc}, and by CMB-S4 to  $|\Delta N_{\rm eff}| \lesssim 0.06$~\cite{Abazajian:2019eic}.

For the exemplary flavor-blind model considered in this work, there is a small preference to heat the $\nu$ sector. The associated temperature evolution $T_\nu/T_\gamma$ explored in Sec.~\ref{sec:evo_temp} therefore translates into an elevated value of  $N_{\rm eff}$.
In the left panel of Fig.~\ref{fig:Neff}, with  benchmark masses $m_\phi = 1 , 5,10\,{\rm MeV}$ we show the $N_{\rm eff}$ prediction for different choices of $\Lambda_{Z'}$.
We find that the Planck+BAO observations yield a lower bound $\Lambda_{Z'} \geq \mathcal{O}(0.08\text{--}0.001)\,{\rm TeV}$ for $m_\phi= 1\text{--}10\,{\rm MeV}$. The associated sensitivity projections for the Simons Observatory and CMB-S4 in the same DM mass range are  $\Lambda_{Z'} \geq \mathcal{O}(0.1\text{--}0.007)\,{\rm TeV}$ and $\Lambda_{Z'} \geq \mathcal{O}(0.15\text{--}0.01)\,{\rm TeV}$, respectively.
The lower panel in Fig.~\ref{fig:Neff} shows the difference of the $N_{\rm eff}$-prediction without DM-SM scattering from the full result. It demonstrates that  the inclusion of elastic scattering affects the value of $N_{\rm eff}$ only weakly. This is expected as the energy-transfer efficiency of scattering is suppressed, compared to that of annihilation; see also Fig.~\ref{fig:DM_Tnuevo} and the discussion in Sec.~\ref{sec:evo_temp}. As shown by the left lower panel of Fig.~\ref{fig:Neff}, the contribution of elastic scattering is reduced both for  large $\Lambda_{Z'}$, where DM only couples feebly to SM particles, and for small $\Lambda_{Z'}$, where sufficient DM annihilation leads to a small DM abundance after freeze-out. 
We conclude that even with upcoming advances, the effect of scattering will be difficult to probe for symmetric DM\footnote{Those conclusions may differ for asymmetric DM with its pronounced role of elastic scattering processes; we shall explore this in future work.}. 

In the right panel of Fig.~\ref{fig:Neff} we show the value of $N_{\rm eff}$ as a function of $m_\phi$ by imposing a thermal $\phi$ relic density of $Y = Y_{\rm DM},\ 0.1 Y_{\rm DM}$ and  $0.01Y_{\rm DM}$ as labeled. Apparently, larger DM-SM interactions reduce the final $\phi$ abundance, but enlarge $N_{\rm eff}$ at CMB.
We observe that Planck+BAO can set a lower limit on $m_\phi \gtrsim 7\,{\rm MeV}$ on thermal DM if the branching ratio of DM annihilation into electrons to that into neutrinos is approximately equipartitioned as by our assumptions.
For $m_\phi > 10\,{\rm MeV}$, DM annihilation decouples just after or even before SM neutrino decoupling to have $Y = Y_{\rm DM}$, thus the effect on $N_{\rm eff}$ is small. Therefore, for larger DM masses, we couple into earlier works~\cite{Escudero:2018mvt} with the conclusion that future CMB-S4 can probe $m_\phi < 15\,{\rm MeV}$. For comparison, we also show the $Y = Y_{\rm DM}$ curve for the annihilation-only case (black dash-dotted line). In terms of the lower limit on $m_\phi$, a $0.6\,{\rm MeV}$ shift from the full case is observed, highlighting again the importance of taking into account the elastic scattering processes.

At last, we mention that for a complex scalar DM our bounds are slightly stronger than those in the previous work. For instance, Ref.~\cite{Sabti:2019mhn} assumes constant annihilation cross sections $\langle \sigma_{{\rm ann.},\,e} v \rangle  = \langle \sigma_{{\rm ann.}, \,\nu} v \rangle  =1.5\times 10^{-26}$\,cm$^3/s$~\cite{Sabti:2019mhn} and obtains $m_\phi \gtrsim 4.5\,{\rm MeV}$, while Ref.~\cite{Depta:2019lbe} concludes $m_\phi \gtrsim 5.4\,{\rm MeV}$ with  $\langle \sigma_{{\rm ann.}, \,e} v \rangle = \langle \sigma_{{\rm ann.},\, \nu} v \rangle  = 4\times 10^{-26}$\,cm$^3/s$. This is mostly because the observed $\Omega_{\rm DM}$ for MeV DM actually requires  larger DM annihilation cross sections than the values adopted in these references, especially before DM-induced energy transfer eventually decouples in the  $p$-wave case. This will be studied in more detail in our follow up work for both $s$- and $p$-wave cases~\cite{companion};  we recall though that $s$-wave DM annihilation into the EM sector is severely constrained from indirect searches.
Our result also improves on previous constraints in the literature based on approximations in DM-SM interactions~\cite{Wilkinson:2016gsy,Boehm:2020wbt} and previous analytical calculations of CMB $N_{\rm eff}$ bound on annihilation of $\mathcal{O}({\rm MeV})$ DM~\cite{Depta:2019lbe}.

\section{Conclusion}
\label{sec:conclusions}

The possibility of DM below the GeV scale has been the center of much attention in recent years. Absent from the literature is a treatment that allows to make precision predictions of both the DM relic abundance and $N_{\rm eff}$ for MeV-scale DM that annihilates into electron (photons) and/or neutrinos with arbitrary branching. The reason is that its abundance is set around neutrino-decoupling and electron annihilation and one must track three coupled sectors across a great dynamical range.

In this work, we lay out a formalism that makes the solution of this problem amenable to ready numerical integration. By taking a series of small but essential approximations we are able to factor out the chemical potentials of neutrinos and DM from the associated distribution functions. Together with a suitable algebraic representation of statistical factors that makes detailed balancing manifest and numerically robust, the collision terms of annihilation and scattering can be solved and the Boltzmann equations integrated. The inclusion of elastic scattering processes is a first in this context and it allows to account for scattering-mediated energy exchange between the various sectors through it.

We test our framework using as an example a flavor-blind vector mediated scalar DM model, with equal couplings to each neutrino flavor and to electrons featuring  $p$-wave annihilation. Such democratic partitioning among the sectors is natural for dark sector particles coupled to the SM lepton ${\rm SU}(2)_L$ doublet. For the purpose of this paper, we take a coupling to lepton number, including both left- and right-handed chiral electrons. Because of this flavor-blindness and spin-statistical factors, the annihilation of $\phi$ heats both sectors almost equally, with a small preference for neutrinos. It may hence be considered a relatively ``safe'' representative of a thermal MeV DM candidate. We obtain the modified evolution of $T_\nu$ and in consequence  $N_{\rm eff}$, and ascertain that elastic scattering plays a subleading role in the prediction of this important observable in such scenario. However, accounting for elastic scattering becomes central if one wishes to track the DM temperature. Its value is bracketed by $T_\nu$ and $T_\gamma$ before its final kinetic decoupling, and enters the DM abundance calculation when kinetic equilibrium is not manifestly assumed. We show that turning on/off the elastic energy transfer to the dark sector affects the DM relic cross section prediction by a factor of a few. In addition, the resulting $N_{\rm eff}$ constraint on the DM mass is also shifted by $0.6\,{\rm MeV}$. In the considered model, using the current CMB constraint, we obtain as minimal thermal scalar DM mass~7~MeV as being currently allowed by observations.

In future work we will employ our established framework to obtain precise predictions on annihilation cross sections of thermal MeV DM candidates with varying branching ratios into EM and $\nu$-sectors, varying spin and varying velocity dependencies, and, more generally, study the principal cosmological viability of light DM candidates today and with upcoming observations. 

\vspace{.3cm}

\paragraph*{Acknowledgments}
 XC and JP are supported by the FWF research group FG01. JLK is supported by the U.S.~National Science Foundation (NSF) Theoretical Physics Program, Grant PHY-1915005. We acknowledge the use of computer packages for algebraic calculations~\cite{Mertig:1990an,Shtabovenko:2016sxi}.
  
\appendix 

\section{Relations between (\boldmath$\rho_i$,\,$n_i$) and (\boldmath$T_i$,\,$\tmu_i$)}
\label{sec:conversion}

With the distribution functions  parametrized by the temperature and chemical potential alone, we may express $n$ and $\rho$ in terms of $T$ and $\mu$, and vice verse.
\paragraph*{EM sector.}  The total energy density in the EM sector may be written as
\begin{equation}
\rho_{\rm EM}  = {\pi^2 \over 30 }g_{\rm EM}(T_\gamma) T_\gamma^4\,, 
\end{equation}
where the effective degrees of freedom of the respective sector, $g_{\rm EM}$, evolves from $(2+4\times 7/8)$ to $2$ during the decoupling of electrons. Using the known function $g_{\rm EM}(T_\gamma)$, one obtains the value of $T_\gamma$ from~$\rho_{\rm EM}$.

\paragraph*{Neutrino sector.} For the massless neutrinos described by Eq.~\eqref{eq:dist_nu}, in a first order expansion in $\tmu_\nu$ one has
\begin{eqnarray} 
  \rho_\nu  &\simeq & {3g_\nu  \,{7\pi^2\over 240} T_\nu^4} \left(  1+ \tmu_\nu \,{540\zeta(3)\over 7\pi^4} \right)\,, \label{eq:nuenergyden}  \\
  n_\nu  &\simeq & {3g_\nu  \,{3\zeta(3)\over 4\pi^2} T_\nu^3} \left( 1+  \tmu_\nu \,{\pi^2 \over 9\zeta(3)} \right) \,,
\end{eqnarray}
where we choose $g_\nu =1$ so it does not count anti-neutrino, and the prefactor, 3, gives the number of SM generations. The ratio of $\rho_\nu^3/n_\nu^4$ is a function of $\tmu_\nu $ only. The final value of $N_{\rm eff}$ is then decided by both, $T_\nu/T_\gamma$ and $\tmu_\nu$. 

\paragraph*{Dark sector.} As $\tmu_\phi$ only enters as a pre-factor, it is easier to first obtain the relation between the average energy $\rho_\phi/n_\phi$ and $T_\phi$ numerically. For our purposes we use the approximation
\begin{equation}
    {\rho_\phi \over n_\phi} \simeq {(3 T_\phi+m_\phi) + \sqrt{(2c_r T_\phi -3 T_\phi)^2  + m_\phi^2}   \over 2 }\,.
\end{equation}
In the relativistic limit, this becomes ${\rho_\phi/ n_\phi} = c_r T_\phi$ with $c_r \equiv 2.701$ (3.151) for bosons (fermions). In turn, in the non-relativistic limit one has ${\rho_\phi/ n_\phi} = m_\phi+ 3T_\phi/2$.

\section{Collision Terms} 
\label{app:collision}

In our convention, $g_i$ counts the (non-identical) {\it particle} degrees of freedom. Hence, we take $g_\phi =1$ for the complex scalar DM particle, $g_e =2$ for the electron, as well as $g_\nu =1$ for each neutrino flavor, distinguishing a chiral neutrino from its antiparticle.
Consequently, each interaction rate below is expressed with respect to its effect on the number/energy density of the particle component only (not including its anti-particle). 

With the three SM generations, the branching ratio of MeV-scale DM annihilation in our flavor-blind setup in the limit of $m_e =0$ is $g_e: 3 g_\nu = 2:3 $. In our benchmark model, DM annihilation hence preferentially heats up the neutrino sector.

\subsection{Interactions within the SM}

We start with the process $\nu\nu \leftrightarrow ee$, for which the number-changing rate and energy-exchange rate are expressed by
\begin{align}
   \left. \dfrac{\delta n}{\delta t}\right|_{\nu\nu \leftrightarrow ee} &= \gamma^{(0)}_{\nu\nu \leftrightarrow ee}  + \beta_{\nu\nu \leftrightarrow ee}^{(1)}  \gamma^{(1)}_{\nu\nu \leftrightarrow ee}\, , \\
   \left. \dfrac{\delta \rho}{\delta t}\right|_{\nu\nu \leftrightarrow ee} &= \zeta^{(0)}_{\nu\nu \leftrightarrow ee} + \beta_{\nu\nu \leftrightarrow ee}^{(1)} \zeta^{(1)}_{\nu\nu \leftrightarrow ee}\,,
\end{align}
 with $\beta_{\nu\nu \leftrightarrow ee}^{(1)} = 2\tmu_\nu$.
 The corresponding $\gamma$- and $\zeta$-collision terms are 
\begin{align*}
\gamma^{(0),(1)}_{\nu\nu \leftrightarrow ee} &= \dfrac{g_\nu^2}{(2\pi)^4} \int \dfrac{ds dE_+ dE_-}{2} \, f_\nu^{\rm eq} f_\nu^{\rm eq, (1)} \sigma_{\nu\nu \rightarrow ee} \mathcal{F}_{12} \nonumber \\
&\times  \left[ (1-\Delta_{\rm ann.}) + \Delta_{\rm ann.} (1- \beta_{\rm ann.})\right]\,,  \\
\zeta^{(0),(1)}_{\nu\nu \leftrightarrow ee} &= \dfrac{g_\nu^2}{(2\pi)^4} \int \dfrac{ds dE_+ dE_-}{2} \, f_\nu^{\rm eq} f_\nu^{\rm eq, (1)} \sigma_{\nu\nu \rightarrow ee} \mathcal{F}_{12}  \nonumber \\
&\times  E_+\left[ (1-\Delta_{\rm ann.}) + \Delta_{\rm ann.} (1- \beta_{\rm ann.})\right]\,,
\end{align*}
where  the total cross section reads
\begin{align}
     \sigma_{\nu\nu \rightarrow ee} &= \dfrac{G_F^2 \sqrt{s-4m_e^2}}{6\pi \sqrt{s}} \nonumber \\
     &\times \left[ m_e^2 (48 s_W^4 -8 s_W^2 -3) +s (24 s_W^4 -4 s_W^2 +3 )\right]\,,
\end{align}
with $G_F$ being the Fermi constant and $s_W \sim 0.47$ is the sine of the Weinberg angle. Here we have summed up the contributions from the three neutrino generations in the cross section which amounts to neutral current processes for $\nu_{e,\mu,\tau}$ and, additionally, the charged current process for $\nu_e$.
In the Maxwell-Boltzmann approximation and massless electron limit, we obtain
\begin{align*}
     \gamma^{(0)}_{\nu\nu \rightarrow ee} &\rightarrow \dfrac{4G_F^2 (24 s_W^4 -4 s_W^2 +3 )(T_\gamma^8-T_\nu^8)}{\pi^5}\,, \\
     \zeta^{(0)}_{\nu\nu \rightarrow ee} &\rightarrow \dfrac{32 G_F^2 (24 s_W^4 -4 s_W^2 +3 )(T_\gamma^9-T_\nu^9)}{\pi^5}\,,
\end{align*}
in agreement with the result in~\cite{Escudero:2020dfa}.
For $\nu e \leftrightarrow \nu e$,  $\gamma =0 $ because of  particle number conservation. 
The energy-exchange rate can be expressed as
\begin{align}
    \left.\dfrac{\delta \rho}{\delta t} \right|_{\nu e \leftrightarrow \nu e} &= \beta^{(0)}_{\nu e \leftrightarrow \nu e} \zeta^{(0)}_{\nu e \leftrightarrow \nu e}+ \beta^{(1)}_{\nu e \leftrightarrow \nu e} \zeta^{(1)}_{\nu e \leftrightarrow \nu e}\,,
\end{align} 
with $\beta^{(0)}_{\nu e \leftrightarrow \nu e} = e^{\tmu_e}$, $\beta^{(1)}_{\nu e \leftrightarrow \nu e}  = e^{\tmu_e} \tmu_\nu$ and 
\begin{align*}
    \zeta^{(0),(1)}_{\nu e \leftrightarrow \nu e} &= \dfrac{ g_\nu  g_e}{(2\pi)^4} \int dE_1 dE_2 ds dt \, f_\nu^{\rm eq, (1)} f_e^{\rm eq} \dfrac{d\sigma_{\nu e \rightarrow \nu e }}{dt} \\ &\times \mathcal{F}_{12} \langle \Delta_{\rm scatt.} \delta E \rangle\,.
\end{align*}
As an exception, here we have already summed up  the differential cross sections of all $\nu e^\mp \rightarrow \nu e^\mp$, $\bar\nu e^\mp \rightarrow \bar\nu e^\mp$  processes for the three  neutrino generations to result in
\begin{widetext}
\begin{align}
 \sum_{\overset{(-)}{\nu}, e^\pm}  \dfrac{ d\sigma_{\nu e \rightarrow \nu e }}{dt} = \dfrac{G_F^2 \left\lbrace (24s_W^4 -4 s_W^2 +3)\left[2(s-m_e^2)^2+2st +t^2\right] -6m_e^2 t\right\rbrace}{2\pi (s- m_e^2)^2}\,.
\end{align} 
\end{widetext}
In the Maxwell-Boltzmann approximation and in the massless electron limit, the integration can be  carried out  analytically
\begin{align*}
    \sum_{\overset{(-)}{\nu} e^\pm}    \zeta^{(0)}_{\nu e \leftrightarrow \nu e} \rightarrow \dfrac{56 G_F^2 (24 s_W^4 -4 s_W^2 +3) T_\nu^4 T_\gamma^4 (T_\gamma- T_\nu)} {\pi^5}\,,
\end{align*}
which is in agreement with the result derived following the appendix in~\cite{Escudero:2020dfa}.

\subsection{DM-electron Interactions}

We now continue collecting the results for interactions between DM and SM. 
For $ee \leftrightarrow \phi\phi$, we may write 
\begin{align}
  \left.\dfrac{\delta n}{\delta t} \right|_{ee \leftrightarrow \phi\phi} &= \beta_{ee \leftrightarrow \phi\phi} \gamma_{ee \leftrightarrow \phi\phi}\,, \\
    \left.\dfrac{\delta \rho}{\delta t} \right|_{ee \leftrightarrow \phi\phi} &= \beta_{ee \leftrightarrow \phi\phi} \zeta_{ee \leftrightarrow \phi\phi} \,,
\end{align}
with $\beta_{ee \leftrightarrow \phi\phi} = e^{2\tmu_e}$ and 
\begin{align*}
\gamma_{ee \leftrightarrow \phi\phi} &= \dfrac{g_e^2}{(2\pi)^4} \int \dfrac{ds dE_+ dE_-}{2} \, f_e^{\rm eq} f_e^{\rm eq} \sigma_{ee \rightarrow \phi\phi} \mathcal{F}_{12} \\
&\times \left[ (1-\Delta_{\rm ann.}) + \Delta_{\rm ann.} (1- \beta_{\rm ann.})\right] \,,\\
\zeta_{ee \leftrightarrow \phi\phi} &= \dfrac{g_e^2}{(2\pi)^4} \int\dfrac{ds dE_+ dE_-}{2} \, f_e^{\rm eq} f_e^{\rm eq} \sigma_{ee \rightarrow \phi\phi} \mathcal{F}_{12} \\
&\times E_+ \left[ (1-\Delta_{\rm ann.}) + \Delta_{\rm ann.} (1- \beta_{\rm ann.})\right] \,.
\end{align*}
In the considered exemplary model, the cross section is given by
\begin{align}
  \sigma_{ee \rightarrow \phi\phi} = \dfrac{(s-4m_\phi^2)^{3/2} (s+2m_e^2)}{48\pi s \sqrt{s-4m_e^2} \Lambda_{Z'}^4}\,.
\end{align}
In the Maxwell-Boltzmann approximation and for $m_\phi = m_e = 0$, we obtain
\begin{align*}
    \gamma_{ee \leftrightarrow \phi\phi} &\rightarrow g_e^2\dfrac{(T_\phi^8-T_\gamma^8)}{2\pi^5 \Lambda_{Z'}^4}\,, \quad
     \zeta_{ee \leftrightarrow \phi\phi} \rightarrow g_e^2  \dfrac{4(T_\phi^9-T_\gamma^9)}{\pi^5 \Lambda_{Z'}^4}\,.
\end{align*} 

Turning to the elastic scattering channel $\phi e \leftrightarrow \phi e$, we have
\begin{align}
    \left.\dfrac{\delta \rho}{\delta t} \right|_{\phi e \leftrightarrow \phi e} &= \beta_{\phi e \leftrightarrow \phi e}  \zeta_{\phi e \leftrightarrow \phi e} \,,
\end{align}
with $\beta_{\phi e \leftrightarrow \phi e}  = e^{\tmu_\phi + \tmu_e}$ and 
\begin{align*}
\zeta_{\phi e \leftrightarrow \phi e} &= \dfrac{(2g_\phi)(2 g_e) }{(2\pi)^4} \int dE_1  dE_2  ds  dt \, f_\phi^{\rm eq}  f_e^{\rm eq} \dfrac{d\sigma_{\phi e \rightarrow \phi e}}{dt} \nonumber \\&\times\mathcal{F}_{12} \langle \Delta_{\rm scatt.}  \delta E \rangle\,,
\end{align*}
where $g_\phi =1$ as by our conventions. The differential cross section for the $Z'$ mediated model and after averaging over all initial states  reads
\begin{align}
\dfrac{d\sigma_{\phi e \rightarrow \phi e}}{dt} = \dfrac{(m_e^2 +m_\phi^2 -s)^2 + t (s-m_e^2)}{4\pi \Lambda_{Z'}^4 [ m_e^4 - 2 m_e^2 (m_\phi^2 +s) +(m_\phi^2 -s )^2 ]}\,.
\end{align}
In the Maxwell-Boltzmann approximation and for $m_\phi = m_e = 0$, the collision term reduces to 
\begin{align}
\zeta_{\phi e \leftrightarrow \phi e}  \rightarrow (g_\phi) (g_e) \dfrac{4  T_\phi^4 T_\gamma^4  ( T_\phi - T_\gamma)}{\pi^5 \Lambda_{Z'}^4}\,,
\end{align}
where we have not taken into account the contributions from the antiparticles, $\phi^*$ and $e^+$, as suggested by the pre-factors. 

\subsection{DM-neutrino Interactions}

Turning to the interaction with neutrinos, for the annihilation $\nu \nu \leftrightarrow \phi\phi$ we again write 
\begin{align}
  \left.\dfrac{\delta n}{\delta t} \right|_{\nu\nu \leftrightarrow \phi\phi} &= \gamma^{(0)}_{\nu\nu \leftrightarrow \phi\phi} +\beta^{(1)}_{\nu\nu \leftrightarrow \phi\phi} \gamma^{(1)}_{\nu\nu \leftrightarrow \phi\phi} \,, \\
    \left.\dfrac{\delta \rho}{\delta t} \right|_{\nu\nu \leftrightarrow \phi\phi} &= \zeta^{(0)}_{\nu\nu \leftrightarrow \phi\phi}+\beta^{(1)}_{\nu\nu \leftrightarrow \phi\phi}  \zeta^{(1)}_{\nu\nu \leftrightarrow \phi\phi} \,,
\end{align}
with $\beta^{(1)}_{\nu\nu \leftrightarrow \phi\phi}  = 2 \tmu_\nu$ and 
\begin{align*}
\gamma^{(0),(1)}_{\nu\nu \leftrightarrow \phi\phi} (T) &= \dfrac{g_\nu^2}{(2\pi)^4} \int \dfrac{dsdE_+ dE_-}{2} \, f_\nu^{\rm eq} f_\nu^{\rm eq, (1)} \sigma_{\nu\nu \rightarrow \phi\phi} \mathcal{F}_{12} \\&\times \left[ (1-\Delta_{\rm ann.}) + \Delta_{\rm ann.} (1- \beta_{\rm ann.})\right]  \,,\\
\zeta^{(0),(1)}_{\nu\nu \leftrightarrow \phi\phi} (T) &= \dfrac{g_\nu^2}{(2\pi)^4} \int \dfrac{dsdE_+ dE_-}{2} \, f_\nu^{\rm eq} f_\nu^{\rm eq, (1)} \sigma_{\nu\nu \rightarrow \phi\phi} \mathcal{F}_{12} \\&\times E_+ \left[ (1-\Delta_{\rm ann.}) + \Delta_{\rm ann.} (1- \beta_{\rm ann.})\right] \,.
\end{align*}
For the chosen model, the cross section is given by
\begin{align}
  \sigma_{\nu\nu \rightarrow \phi\phi} = 3\,\dfrac{(s-4m_\phi^2)^{3/2} }{24\pi \sqrt{s} \Lambda_{Z'}^4}\,,
\end{align}
where the factor of~3 is for three neutrino generations. For each generation, the result  is approximately twice that for electrons above, since for electrons one has to average over non-contributing initial states,  $(e^-_L e^+_L)$ and  $(e^-_R e^+_R)$.

In the Maxwell-Boltzmann approximation and for $m_\phi = 0$ we find
\begin{align*}
    \gamma^{(0)}_{\nu\nu \leftrightarrow \phi\phi}  &\rightarrow (3 g_\nu^2) \dfrac{(T_\phi^8 -T_\nu^8)}{\pi^5 \Lambda_{Z'}^4}\,, \\
     \zeta^{(0)}_{\nu\nu \leftrightarrow \phi\phi} &\rightarrow (3 g_\nu^2)  \dfrac{8(T_\phi^9-T_\nu^9)}{\pi^5 \Lambda_{Z'}^4}\,.
\end{align*}

Finally, for the elastic scattering $\phi \nu \leftrightarrow \phi \nu$, we write 
\begin{align}
    \left.\dfrac{\delta \rho}{\delta t} \right|_{\phi \nu \leftrightarrow \phi \nu} &= \beta^{(0)}_{\phi \nu \leftrightarrow \phi \nu}  \zeta^{(0)}_{\phi \nu \leftrightarrow \phi \nu} + \beta^{(1)}_{\phi \nu \leftrightarrow \phi \nu}  \zeta^{(1)}_{\phi \nu \leftrightarrow \phi \nu}\,,
\end{align} 
with $\beta^{(0)}_{\phi \nu \leftrightarrow \phi \nu} = e^{\tmu_\phi}$, $\beta^{(1)}_{\phi \nu \leftrightarrow \phi \nu} = e^{\tmu_\phi} \tmu_\nu$ and 
\begin{align*}
    \zeta^{(0),(1)}_{\phi \nu \leftrightarrow \phi \nu} &= \dfrac{g_\phi g_\nu }{(2\pi)^4} \int dE_1 dE_2 ds dt \, f_\phi^{\rm eq} f_\nu^{\rm eq, (1)} \dfrac{d\sigma_{\phi \nu \rightarrow \phi \nu }}{dt} \\&\times \mathcal{F}_{12} \langle \Delta_{\rm scatt.} \delta E \rangle\,.
\end{align*}
The differential cross section in the exemplary model reads
\begin{align}
   \dfrac{ d\sigma_{\phi \nu \rightarrow \phi \nu }}{dt} = 3 \, \dfrac{[(m_\phi^2 -s)^2 +st ]}{4 \pi \Lambda_{Z'}^4 (m_\phi^2 -s)^2}\,,
\end{align} 
and it coincides with the  electron case for $m_e =0$ (when the factor of 3 is dropped).
In the Maxwell-Boltzmann  massless DM approximation, we obtain
\begin{align}
      \zeta^{(0)}_{\phi \nu \leftrightarrow \phi \nu }  \rightarrow (g_\phi)(3 g_\nu)  \dfrac{4 T_\phi^4 T_\nu^4 (T_\phi - T_\nu)} {\pi^5 \Lambda_{Z'}^4}\,.
\end{align}
Here we have accounted for the three neutrino generations, but, as usual, not for the contribution from $\phi^*$ and~$\bar{\nu}$.

\bibliography{refs}

\end{document}